%% file: Manuscript.tex
\documentclass[12pt]{article}
\usepackage{amsfonts}
\usepackage{amsmath, amsthm}
\usepackage{epsfig}
\usepackage{algorithm}
\usepackage{algorithmic}
\usepackage{epic}
\usepackage{threeparttable}

\usepackage[a4paper]{geometry}
\usepackage{enumerate}
\usepackage{amsmath,verbatim,color,amssymb,epsfig}
\usepackage{bm}
\usepackage{ragged2e}
\usepackage{amsfonts}
\usepackage{epsfig}
\usepackage{changepage}
\usepackage{multirow}
\usepackage{graphicx}
\usepackage{array}
\usepackage[table]{xcolor}
\definecolor{Gray}{gray}{.80}
\usepackage{lscape}
\usepackage{enumitem}

\usepackage[round,semicolon,authoryear]{natbib}
\usepackage[normalem]{ulem}
\usepackage[colorlinks=true,urlcolor=blue,citecolor=purple,linkcolor=blue,bookmarks=true]{hyperref}
\usepackage{caption}
\usepackage{pdfsync}
\usepackage{setspace}

\input{def}

\usepackage{booktabs} 
\usepackage{cleveref} 
\usepackage{caption}  
\usepackage{tabularx} 
\usepackage{adjustbox}
\usepackage{longtable}

\author{Xu}
\date{November 2024}

\begin{document}
\doublespacing   
\baselineskip=12pt
\begin{center}
{\Large \bf A staggered seamless dose-optimization design for co-developing monotherapy and combination therapy}
\end{center}
 \vspace{1mm}
 \begin{center}
 {\bf Masahiro Kojima$^{1}$, Kentaro Takeda$^{2*}$, Ying Yuan$^{3}$}\\
 \vspace{0.5cm}
 \noindent $^{1}$Chuo University, Bunkyo-ku, Tokyo, Japan.\\\vspace{1mm} $^{2}$Astellas Pharma Global Development Inc., Northbrook, IL, USA. \\\vspace{1mm} $^{3}$The University of Texas MD Anderson Cancer Center, Houston, Texas, USA
 \\
 \vspace{1mm} 
 $^*$Authors for correspondence: kentaro.takeda@astellas.com
 \end{center}
 \vspace{2mm}

\baselineskip=24pt

\noindent \emph{\textbf{Abstract}}: Contemporary oncology drug development increasingly requires efficient dose-optimization strategies that evaluate monotherapy (Mono) and combination therapy (Combo) while balancing activity, efficacy, and tolerability. We propose a staggered seamless phase I/II design for settings in which a novel agent is evaluated alone and in combination with an established therapy. In phase I, Mono dose finding begins first, and Combo subtrials can be opened adaptively once a prespecified combination-initiation signal based on early clinical or biological information is observed. Dose assignment uses a model-assisted rule based on toxicity and early activity, with backfilling at tolerable and potentially promising regimens. At the end of phase I, two candidate regimens are selected from the evaluated Mono and Combo regimens using an efficacy-toxicity utility based on accumulated toxicity and treatment-response data. Phase II seamlessly carries forward patients treated at the selected regimens, enrolls additional patients as needed, and applies Bayesian futility and efficacy stopping boundaries to identify a final recommended optimal biological dose (OBD). Simulation studies showed that the proposed design shortened phase I trial duration relative to the comparator designs while maintaining competitive OBD-selection performance and acceptable safety. The seamless phase II component further reduced the need for additional enrollment and supported efficient final OBD selection.

\vspace{0.5cm}

\noindent \emph{\textbf{keywords}}: dose-optimization; monotherapy; combination therapy; model-assisted design; backfill.


\section{Introduction}\label{sec_intro}

The emergence of anticancer agents with distinct mechanisms of action has expanded therapeutic options and accelerated the development of rational combination strategies designed to enhance anticancer activity, improve the durability of clinical benefit, and overcome intrinsic and acquired resistance~\citep{coleman2026biology}. Consistent with this biological and clinical rationale, phase~I oncology trials have increasingly featured multi-agent combination studies, highlighting the growing importance of combination development in contemporary early-phase oncology~\citep{kelly2024changing}. These trends create a need for trial designs that can characterize monotherapy (Mono) and evaluate combination therapy (Combo) within an integrated development framework.

Under the conventional fully sequential Mono--Combo paradigm, the evaluation of potentially promising combination regimens can be delayed because the Combo component is typically initiated only after Mono dose finding has been completed. To expedite the development of both monotherapy and combination therapy, it may be desirable to initiate the Combo component before the Mono dose escalation has fully concluded, provided that accumulating Mono data show an acceptable safety profile together with sufficient clinical or biological evidence to justify combination evaluation. Such a strategy allows earlier evaluation of well-justified combination regimens while the Mono dose and schedule are still being refined.

For example, in the first-in-patient phase~1b dose-escalation and dose-expansion trial of zandelisib (ME-401) administered as Mono or in combination with rituximab~\citep{pagel2022zandelisib}, seven dose levels were originally planned for dose escalation, but only three dose levels (60, 120, and 180 mg) were evaluated. Because no dose-limiting toxicities were observed and safety and preliminary antitumor activity were similar across the tested doses, further escalation solely to define the maximum tolerated dose was not pursued. Instead, the lowest tested dose, 60 mg, was selected as the minimum biologically effective dose and recommended phase~II dose. Subsequent evaluation then proceeded with the 60 mg dose, including schedule optimization of zandelisib monotherapy and evaluation of zandelisib in combination with rituximab.

As illustrated by the zandelisib example, early Mono data can identify a biologically justified dose before Mono dose exploration has fully concluded. This supports the need for a trial design framework in which Combo evaluation can be opened adaptively once accumulating Mono data provide credible evidence of activity and acceptable safety.

More broadly, early-phase dose development increasingly aims to justify dose and schedule through an explicit benefit--risk assessment of efficacy and tolerability (e.g., selection of an optimal biological dose, OBD), rather than relying solely on identification of the maximum tolerated dose (MTD), as highlighted in initiatives such as the U.S. FDA's Project Optimus~\citep{fda_project_optimus_2024}. Achieving this goal typically requires evidence across multiple doses and, in many settings, efficacy assessment takes longer to mature than toxicity. As a result, fully sequential evaluation of Mono doses followed by combination regimens can be time-consuming, particularly when many candidate regimens are under consideration.

Several approaches have been considered for integrating monotherapy and combination therapy development. \citet{zhou2024oncology} reviewed practical development strategies for oncology combination dose finding, including sequential escalation after monotherapy and parallel (staggered) escalation with monotherapy; their work focused on development strategies and considerations rather than proposing a specific statistical dose-assignment method. More recently, \citet{daniells2025seamless} proposed a model-based phase~I design for parallel monotherapy and combination dose escalation in a setting with a fixed-dose backbone agent. Their design shares toxicity information between the monotherapy and combination components to identify the monotherapy MTD and the maximum tolerated combination, while efficacy or early activity is not incorporated into the dose-assignment decisions. In later-stage development, Bayesian randomized phase~II designs have been proposed to integrate combination dose optimization with assessment of the contribution of individual components~\citep{chi2025coca,chi2026bop2}. These studies address complementary aspects of monotherapy and combination therapy development. The present work focuses on a staggered phase~I/II setting in which both agents may have multiple candidate dose levels, Combo subtrials are opened adaptively using accumulating Mono toxicity and early-activity information, and selected regimens are carried forward seamlessly into phase~II evaluation.

In this paper, we propose an accelerated seamless staggered phase~I/II design that enables an earlier transition from Mono to combination evaluation and leverages phase~I data to streamline the subsequent phase~II assessment. Once a prespecified combination-initiation signal is observed in the Mono component, a Combo subtrial can be opened adaptively, allowing enrollment into the combination component before Mono dose exploration is completed, while making use of accumulating Mono information. This signal is intended to be broader than the formal efficacy endpoint used for dose optimization and may be based on early antitumor activity, PK/PD information, target engagement, or biomarker evidence, depending on the clinical and biological context. Importantly, this signal is used for early phase~I decisions, including dose-finding decisions and initiation of Combo subtrials, whereas final candidate selection and phase~II evaluation are based on formal tumor response and toxicity data. After the Mono/Combo component, we select the two candidate regimens with the highest utility from among all evaluated Mono and Combo regimens, so the selected pair may consist of monotherapy, combination therapy, or both. These two selected regimens then enter the phase~II component for comparative evaluation with prespecified futility and efficacy monitoring. A key feature of our proposal is that the phase~II evaluation is also seamless: patients accrued at each selected candidate regimen in phase~I are carried forward and incorporated into the phase~II analysis. By leveraging phase~I data at the two selected candidate regimens, the proposed phase~II procedure can reduce the number of additional patients required and shorten the overall evaluation time, while maintaining rigorous futility/efficacy monitoring and enabling an efficient final recommendation. Among the candidate regimens declared promising in phase~II, the one with the greater utility is selected as the final recommended OBD. If neither candidate regimen is declared promising, no OBD is selected. Overall, the proposed approach is seamless both in moving from Mono to Combo within phase~I and in transitioning from phase~I to phase~II, with the goal of improving efficiency without compromising decision quality.

The remainder of this paper is organized as follows. Section~2 presents motivating examples to provide intuition for the proposed approach. Section~3 describes the proposed trial design in detail. Section~4 evaluates operating characteristics through simulation studies. Section~5 concludes with a discussion of limitations and future directions.

\section{Illustrative example of the staggered Phase I dose-assignment design}\label{motivation}

To provide intuition for the proposed staggered strategy before presenting the formal methodology, we first introduce a simple illustrative example of the Phase~I dose-assignment process. This example focuses on how information accumulated in the Mono component can trigger the adaptive opening of Combo subtrials while Mono dose exploration is still ongoing.

\begin{figure}[tbp]\centering
\includegraphics[width=10cm]{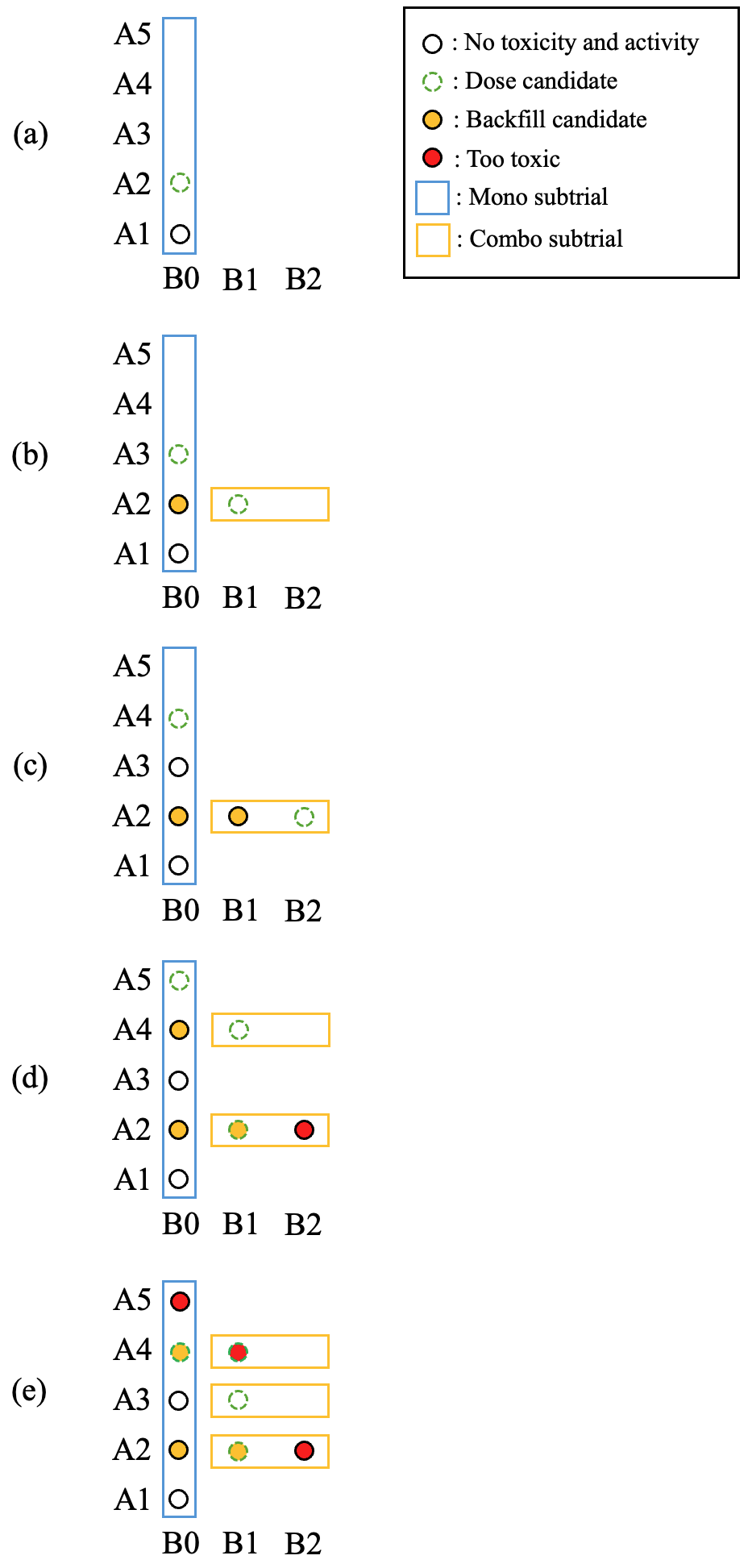}
\caption{Conceptual illustration of the staggered Phase I dose-assignment workflow from Mono to Combo.}
\label{fig:Motivating_Example}
\end{figure}

Figure~\ref{fig:Motivating_Example} provides a conceptual illustration of this staggered Mono-to-Combo workflow. In this example, agent $A$, the primary investigational agent, has five dose levels, $A_1, \ldots, A_5$. Agent $B$ is an existing drug, and the objective is to explore candidate regimens involving agent $A$, either alone or in combination with agent $B$. Let $B_0$ denote no administration of agent $B$, so that regimens involving $B_0$ correspond to Mono with agent $A$. For agent $B$, two active dose levels, $B_1$ and $B_2$, are considered.

For simplicity, this illustrative example represents the combination-initiation signal by early antitumor activity observed in the Mono component. In practice, this signal may also be based on PK/PD information, target engagement, or biomarker evidence, and should be prespecified in the protocol. In the description below, dose candidates denote regimens eligible to receive the next dose-escalation cohort. When multiple dose candidates are available and the corresponding dose-escalation cohorts have not reached their planned sizes, patients may be randomized among the dose candidates. Backfill candidates denote tolerable and potentially promising regimens at which additional patients may be enrolled when all active dose-escalation cohorts are full and their patients are still under evaluation.

\begin{itemize}
\item[(a)] The trial starts at $A_1B_0$. Because neither toxicity nor a combination-initiation signal is observed at $A_1B_0$, the next Mono dose candidate is $A_2B_0$.

\item[(b)] At $A_2B_0$, a combination-initiation signal is observed with acceptable toxicity. The next Mono dose candidate is $A_3B_0$, and the corresponding Combo subtrial at the $A_2$ level is opened. The first Combo dose candidate in this subtrial is $A_2B_1$. Thus, patients may be randomized between $A_3B_0$ and $A_2B_1$ when both dose candidates are available. Because activity has been observed at $A_2B_0$ with acceptable toxicity, $A_2B_0$ also becomes a backfill candidate.

\item[(c)] At $A_3B_0$, no combination-initiation signal is observed. The next Mono dose candidate is $A_4B_0$, and the corresponding Combo subtrial at the $A_3$ level is not opened. Meanwhile, in the previously opened $A_2$ Combo subtrial, the dose candidate moves from $A_2B_1$ to $A_2B_2$ according to the Combo dose-assignment rule. Thus, patients may be randomized between $A_4B_0$ and $A_2B_2$ when both dose candidates are available. The regimens $A_2B_0$ and $A_2B_1$ are retained as backfill candidates.

\item[(d)] At $A_4B_0$, a combination-initiation signal is observed with acceptable toxicity. The next Mono dose candidate is $A_5B_0$, and the corresponding Combo subtrial at the $A_4$ level is opened, with $A_4B_1$ as the first Combo dose candidate. In the $A_2$ Combo subtrial, toxicity is observed at $A_2B_2$, so further escalation within that subtrial is not pursued and $A_2B_1$ remains the lower-dose candidate for further evaluation. At this stage, patients may be randomized among available active dose candidates, including $A_5B_0$, $A_2B_1$, and $A_4B_1$. The regimens $A_2B_0$, $A_2B_1$, and $A_4B_0$ may serve as backfill candidates.

\item[(e)] At $A_5B_0$, toxicity is observed, and the next Mono dose candidate is $A_4B_0$ (de-escalation). Toxicity is also observed at $A_4B_1$. Because $A_4B_1$ is the lowest active dose of agent $B$ within the $A_4$ Combo subtrial, the next candidate regimen within that subtrial remains $A_4B_1$; at the same time, the lower-dose Combo subtrial at the $A_3$ level is opened, with $A_3B_1$ as the first Combo dose candidate. At this stage, patients may be randomized among available active dose candidates, including $A_4B_0$, $A_2B_1$, $A_3B_1$, and $A_4B_1$. Backfilling may be performed at selected tolerable and potentially promising regimens, such as $A_2B_0$, $A_2B_1$, and $A_4B_0$.

\end{itemize}

To simplify the presentation, the panels in this example are displayed as if dose-assignment decisions occur in a synchronized sequence across subtrials. In the proposed design, however, dose assignment can be carried out independently within each subtrial as patient outcomes become available.

\section{Methodology}\label{methodology}
We consider a seamless staggered phase~I/II trial. Phase~I consists of two dose-finding components: the Mono component and the Combo component.

\subsection{Phase~I: Mono and Combo dose-escalation design}
Let agent $A$ be a novel investigational anticancer agent with $L$ dose levels, denoted by $A_1, A_2, \ldots, A_L$, and let agent $B$ be an established standard-of-care therapy with $M+1$ dose levels, denoted by $B_0, B_1, \ldots, B_M$. Here, $B_0$ represents no administration of agent $B$; for example, the regimen $A_1B_0$ corresponds to Mono with agent $A$ at dose level $A_1$. Under this notation, $A_lB_m$ provides a unified representation of both Mono with agent $A$ ($m=0$) and Combo with agent $B$ ($m \ge 1$). We use the term candidate regimen to refer to any $A_lB_m$, including both monotherapy regimens ($m=0$) and combination regimens ($m\ge1$). We reserve the term combination regimen for regimens with an active dose of agent $B$.

We view all candidate regimens as an $L \times (M+1)$ regimen matrix and, following \citet{Zhang2016-fp} and \citet{Kakizume2024-sr}, partition this matrix into a Mono subtrial and multiple Combo subtrials. The Mono subtrial is defined as
$\mathcal{S}_1=\{A_1B_0, A_2B_0, \ldots, A_LB_0\}$. For each Mono dose level $A_l$ $(l=1,\ldots,L)$, the corresponding Combo subtrial is defined as $\mathcal{S}_{l+1}=\{A_lB_1,\ldots,A_lB_M\}$. Thus, $\mathcal{S}_{l+1}$ denotes the Combo subtrial associated with Mono dose level $A_l$, and it is opened when the corresponding criterion in the Mono or Combo component is met.

Phase~I dose assignment is based on a model-assisted design. In the proposed design, we consider early activity as the efficacy-type information used for real-time Phase~I decision-making. Here, early activity is intended to represent short-term antitumor activity, PK/PD information, target engagement, or biomarker evidence observed within an early treatment cycle, rather than the formal efficacy endpoint used for final dose optimization. This distinction is important because formal efficacy endpoints, such as ORR or tumor response, may require longer follow-up and therefore may delay efficient dose assignment and the initiation of subtrials. In the Mono component, toxicity and early activity jointly determine whether to escalate, stay, de-escalate, and open the corresponding Combo subtrial. Once a Combo subtrial is opened, the same toxicity- and early-activity-based model-assisted decision framework is applied within that subtrial to guide dose assignment along the dose levels of agent~$B$.

To incorporate information from patients with pending outcomes, particularly backfilled patients, we use a likelihood-based pending-outcome approximation for model-assisted designs~\citep{LinYuan2020,Takeda2020TITEBOINET,takeda2025bf}. For endpoint $X\in\{T,S\}$, let $w_{Xlmi}=\Pr(t_{Xlmi}\le u_{Xlmi}\mid y_{Xlmi}=1)$ denote the weight for a patient whose outcome is pending at the decision time. Following the uniform weighting scheme, we assume that the event time conditional on occurrence is uniformly distributed over the assessment window $(0,\tau_X)$, yielding
\[
w_{Xlmi}=\frac{u_{Xlmi}}{\tau_X}.
\]
We define the number of observed events by the decision time and the effective sample size as
\[
\xt_{Xlm}=\sum_{i=1}^{n_{lm}}\tilde y_{Xlmi},
\qquad
\nt_{Xlm}=\sum_{i=1}^{n_{lm}}
\left\{
\gamma_{Xlmi}
+(1-\gamma_{Xlmi})w_{Xlmi}
\right\},
\]
or equivalently, under the uniform weighting scheme,
\[
\nt_{Xlm}=\sum_{i=1}^{n_{lm}}
\left\{
\gamma_{Xlmi}
+(1-\gamma_{Xlmi})\frac{u_{Xlmi}}{\tau_X}
\right\}.
\]
The effective data are denoted by $\Dt_{Xlm}=(\nt_{Xlm},\xt_{Xlm})$, and the corresponding pending-outcome--adjusted estimate is
\[
\pt_{Xlm}=\frac{\xt_{Xlm}}{\nt_{Xlm}}.
\]
Thus, fully ascertained outcomes contribute one patient to the effective sample size, whereas patients with pending outcomes contribute according to the proportion of the assessment window completed.

To describe the dose-assignment rules, we introduce the decision boundaries for toxicity and early activity. Let $\lambda_{T1}$ denote the lower toxicity boundary for escalation, $\lambda_{T2}$ the upper toxicity boundary for de-escalation, and $\lambda_S$ the early-activity boundary. The dose-assignment rules for the proposed staggered Mono/Combo design are given below.

For notational convenience, let
$\mathcal{S}_{l+1}=\{A_lB_1,\ldots,A_lB_M\}$
denote the Combo subtrial associated with Mono dose level $A_l$.

\paragraph{Mono dose-assignment rule.} The Mono subtrial is $\mathcal{S}_{1}=\{A_{1}B_{0},\ldots,A_{L}B_{0}\}$. For the current Mono regimen $A_lB_0$, the dose-assignment decision is made as follows:
\begin{enumerate}[label={(M-\Alph*)}, leftmargin=*]
\item \textit{Low toxicity with an early activity signal}: If $\pt_{Tl0}\le\lambda_{T1}$ and $\pt_{Sl0}\ge\lambda_S$, assign the next Mono dose-escalation cohort to $A_{l+1}B_0$. In addition, if the corresponding Combo subtrial $\mathcal{S}_{l+1}=\{A_lB_1,\ldots,A_lB_M\}$ has not yet been opened, open it and initiate the Combo dose-assignment rule described below in parallel.
\item \textit{Low toxicity without an early activity signal}: If $\pt_{Tl0}\le\lambda_{T1}$ and $\pt_{Sl0}<\lambda_S$, assign the next Mono dose-escalation cohort to $A_{l+1}B_0$.
\item \textit{Acceptable toxicity with an early activity signal}: If $\lambda_{T1}<\pt_{Tl0}\le\lambda_{T2}$ and $\pt_{Sl0}\ge\lambda_S$, assign the next Mono dose-escalation cohort to $A_lB_0$. In addition, if the corresponding Combo subtrial $\mathcal{S}_{l+1}=\{A_lB_1,\ldots,A_lB_M\}$ has not yet been opened, open it and initiate the Combo dose-assignment rule described below in parallel.
\item \textit{Acceptable toxicity without an early activity signal}: If $\lambda_{T1}<\pt_{Tl0}\le\lambda_{T2}$ and $\pt_{Sl0}<\lambda_S$, assign the next Mono dose-escalation cohort to $A_lB_0$.
\item \textit{Excessive toxicity}: If $\pt_{Tl0}>\lambda_{T2}$, assign the next Mono dose-escalation cohort to $A_{l-1}B_0$.
\end{enumerate}

\paragraph{Dose-assignment rule for an opened Combo subtrial.} For an opened Combo subtrial $\mathcal{S}_{l+1}=\{A_lB_1,\ldots,A_lB_M\}$, the initial Combo dose candidate is $A_lB_1$. Dose assignment within this subtrial is governed by the same toxicity- and early-activity-based model-assisted decision framework used in the Mono component, applied along the dose levels of agent~$B$ with $A_l$ fixed. For the current Combo regimen $A_lB_m$, the dose-assignment decision is made as follows:
\begin{enumerate}[label={(C-\Alph*)}, leftmargin=*]
\item \textit{Escalation region}: If the decision rule recommends escalation at $A_lB_m$, assign the next Combo dose-escalation cohort to $A_lB_{m+1}$.
\item \textit{Stay region}: If the decision rule recommends staying at $A_lB_m$, assign the next Combo dose-escalation cohort to $A_lB_m$.
\item \textit{De-escalation region}: If the decision rule recommends de-escalation at $A_lB_m$, assign the next Combo dose-escalation cohort to $A_lB_{m-1}$ when $m\neq1$. When $m=1$, retain $A_lB_1$ as the next Combo dose candidate. If $l>1$ and the lower-dose Combo subtrial $\mathcal{S}_{l}=\{A_{l-1}B_1,\ldots,A_{l-1}B_M\}$ has not yet been opened, open it and initiate its Combo dose-assignment rule in parallel.
\end{enumerate}

Initially, only the Mono subtrial is active. When criterion~(M-A) or (M-C) is met at a Mono regimen $A_lB_0$, the corresponding Combo subtrial $\mathcal{S}_{l+1}$ is opened and its Combo dose-assignment rule is applied in parallel with the ongoing Mono subtrial. If the Mono subtrial is completed without ever meeting criterion~(M-A) or (M-C), the Combo component is not initiated.

\begin{figure}[tbp]
\centering 
\includegraphics[width=15cm]{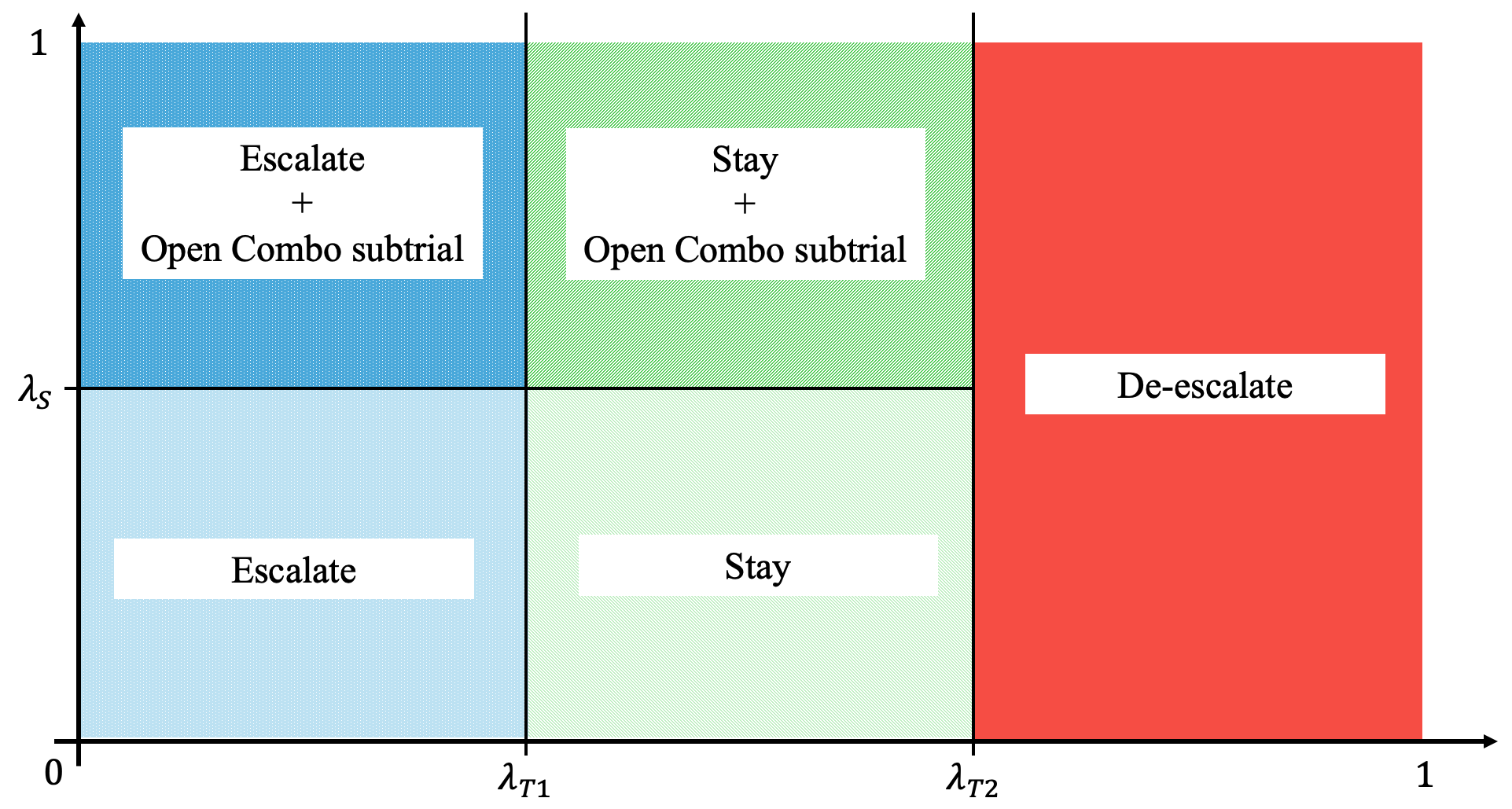} \caption{Visual summary of the Mono dose-assignment rule and the criteria for opening Combo subtrials.} \label{fig:dose_open_example} 
\end{figure}

A key feature of the proposed design is that, as additional Mono dose levels satisfy criterion~(M-A) or (M-C), additional Combo subtrials can be opened, allowing dose assignment to proceed in parallel across the Mono subtrial and the opened Combo subtrials. A visual summary of the Mono dose-assignment rule for opening Combo subtrials is presented in Figure~\ref{fig:dose_open_example}. 

When multiple active dose-escalation cohorts are available across the Mono and opened Combo subtrials, patients are randomized among cohorts that have not reached their planned cohort sizes. To avoid spreading accrual too thinly across many cohorts, a utility-based prioritization rule may be used to favor more promising cohorts.

Enrollment into active dose-escalation cohorts in the Mono and Combo components is prioritized. If a new patient becomes available for assignment when all active dose-escalation cohorts have already reached their prespecified cohort sizes and are awaiting sufficient outcome information for the next dose-assignment decision, the patient is assigned for backfilling to a regimen selected from the set of backfill candidate regimens. The backfill regimen is selected as the regimen that maximizes the early activity--toxicity utility, and patients are enrolled at that regimen. To quantify the early activity--toxicity trade-off for Phase~I backfilling, we use the utility proposed by \citet{Zhou2019-nh, Lin2020-vf}:
\[
U^{(S)}_{lm}=\sum_{u=1}^{U} \psi_{u} q^{(S)}_{lm,u},
\]
where $\psi_{u}$ $(u=1,\ldots,U)$ denotes the utility score for the $u$th possible early activity--toxicity outcome, and $q^{(S)}_{lm,u}$ denotes the corresponding probability at regimen $A_lB_m$. All patients in the dose-escalation cohort must complete the toxicity evaluation period before the next dose assignment is made. In contrast, some backfilled patients may not have completed their evaluation period when the next dose for the dose-escalation cohort is determined. Such patients continue treatment until they complete the evaluation period.

For each subtrial, enrollment is stopped when either (1) the total number of patients in the subtrial reaches the maximum sample size $N_{\textup{maxsub}}$ or (2) the total number of patients treated at the current regimen through dose escalation and backfilling reaches the maximum sample size per regimen, $n_{\textup{stop}}$, and the same regimen is recommended again as the next dose. To avoid assigning patients to regimens with insufficient early activity or excessive toxicity, regimen-elimination criteria are applied before dose assignment.

Using the effective binomial likelihood based on $\Dt_{Xlm}$, we assume independent $\mathrm{Beta}(a_X,b_X)$ priors for $p_{Xlm}$, $X\in\{T,S\}$. The corresponding posterior distribution is
\[
p_{Xlm}\mid\Dt_{Xlm}
\sim
\mathrm{Beta}\left(a_X+\xt_{Xlm},\,b_X+\nt_{Xlm}-\xt_{Xlm}\right),
\]
where $a_X$ and $b_X$ are prespecified hyperparameters. In our implementation, we set $a_X=b_X=1$ for both toxicity and early activity.

Let $\phi_{T0}$ denote the toxicity threshold for elimination and $\phi_{S0}$ the minimum acceptable early-activity probability. Let $C_T$ and $C_S$ denote the corresponding posterior probability cutoffs for toxicity and early-activity elimination, respectively. The regimen-elimination criteria are:
\begin{itemize}
    \item Eliminate regimen $A_lB_m$ and all regimens in the monotone upper-right region $\{(l',m'): l'\ge l,\ m'\ge m\}$ if
    \[
    \textup{Pr}(p_{Tlm}>\phi_{T0}\mid\Dt_{Tlm})>C_T.
    \]
    \item Eliminate regimen $A_lB_m$ for insufficient early activity if
    \[
    \textup{Pr}(p_{Slm}\le\phi_{S0}\mid\Dt_{Slm})>C_S.
    \]
\end{itemize}
The trial is terminated if all candidate regimens are eliminated.

\subsubsection{Selection of Candidate Regimens for Phase~II}\label{selection}
At the end of Phase~I, candidate regimens for Phase~II are selected using accumulated toxicity and treatment response data. The Phase~I ranking procedure can accommodate a prespecified number of candidate regimens; in this paper, we carry forward the two highest-ranked candidate regimens for the subsequent Phase~II procedure and simulation studies. The candidate regimens may include monotherapy, combination therapy, or both. Candidate regimens are selected from an admissible set $\mathcal{D}$, defined below, such that eligible regimens do not exceed the corresponding maximum tolerated regimen within each dose level of agent~$B$.

We assume that the toxicity probability is nondecreasing in the dose levels of both agents. To enforce this monotonicity across all candidate regimens, we first obtain an estimated toxicity probability for every regimen, including those not administered during the trial, using the following beta--binomial model:
$$
x_{Tlm} \mid \pi_{Tlm} \sim \textup{Binomial}(n_{lm}, \pi_{Tlm}), 
\qquad
\pi_{Tlm} \sim \textup{Beta}(0.05, 0.05),
$$
where $n_{lm}$ denotes the number of patients treated at regimen $A_lB_m$ in Phase~I, and $x_{Tlm}$ denotes the corresponding number of observed toxicities. The posterior mean of $\pi_{Tlm}$ is $\hat{\pi}_{Tlm} = (x_{Tlm}+0.05)/(n_{lm}+0.1).$
We then apply bivariate isotonic regression (two-dimensional PAVA) to $\hat{\pi}_{Tlm}$ to obtain the monotone-adjusted toxicity estimates $\bar{p}_{Tlm}$ for each regimen $(A_l,B_m)$.

For each dose level $m$ of agent $B$, the maximum tolerated regimen, denoted by $d_{\textup{MTR},m}$, is selected as the regimen whose estimated toxicity probability is closest to the target toxicity rate $\phi$, that is,
$$
d_{\textup{MTR},m}
=
\underset{l \in \{1,\cdots,L\}}{\operatorname{arg\,min}}
\left|\bar{p}_{Tlm}-\phi\right|,
\qquad (m=0,1,\cdots,M).
$$
The admissible set $\mathcal{D}$ consists of all evaluated, non-eliminated regimens that do not exceed $d_{\textup{MTR},m}$ within each dose level $m$ of agent~$B$. For example, assuming that the relevant regimens have been evaluated and none has been eliminated, in a $3\times2$ regimen matrix, if $\{d_{\textup{MTR},0}, d_{\textup{MTR},1}\}=\{A_3B_0, A_2B_1\},$ then the admissible regimens are $\{A_1B_0, A_2B_0, A_3B_0, A_1B_1, A_2B_1\}.$
For efficacy, we use the observed treatment response probability as the estimated efficacy probability, i.e., $\bar{p}_{Elm}=\hat{p}_{Elm}.$

Based on the estimated toxicity and efficacy probabilities, $\bar{p}_{Tlm}$ and $\bar{p}_{Elm}$, together with the efficacy--toxicity trade-off utility introduced in \citet{takeda2025bf}, the estimated mean utility for regimen $A_lB_m$ at the end of Phase~I is given by
\begin{align}
\bar{U}^{(E)}(A_l,B_m)=\sum_{u=1}^{U}\psi_u \bar{q}^{(E)}_{lm,u},
\qquad (A_l,B_m)\in \mathcal{D},
\end{align}
where $\psi_u~(u=1,\ldots,U)$ denotes the utility score for the $u$th possible efficacy--toxicity outcome, and $\bar{q}^{(E)}_{lm,u}$ denotes the corresponding probability of that outcome at regimen $A_lB_m$. For example, when $U=4$, $\bar{q}^{(E)}_{lm,1}=\bar{p}_{Elm}(1-\bar{p}_{Tlm})$, $\bar{q}^{(E)}_{lm,2}=\bar{p}_{Elm}\bar{p}_{Tlm}$, 
$\bar{q}^{(E)}_{lm,3}=(1-\bar{p}_{Elm})(1-\bar{p}_{Tlm})$, and
$\bar{q}^{(E)}_{lm,4}=(1-\bar{p}_{Elm})\bar{p}_{Tlm}$.
The two candidate regimens with the largest estimated utilities are then selected for Phase~II.

\subsection{Phase~II: Dose Optimization and Efficacy Evaluation}\label{stage2}
In contrast to the early activity signal used for real-time Phase~I decisions, Phase~II evaluates the selected candidate regimens using the formal efficacy endpoint, such as tumor response, together with toxicity. Use of the later efficacy endpoint is operationally feasible in Phase~II because, unlike Phase~I, Phase~II does not require frequent real-time dose-assignment decisions and instead evaluates the selected regimens only at prespecified interim analyses.

In Phase~I, we developed a utility-based strategy for exploring candidate regimens. In Phase~II, building on the futility and efficacy stopping rules studied by \citet{Zhou2017-tv,Xu2025-xq}, we propose a seamless evaluation procedure that directly carries forward the data accrued at the two candidate regimens selected in Phase~I and incorporates them into the Phase~II analysis, thereby reducing additional enrollment and shortening the overall Phase~II evaluation period.

\subsubsection{Stopping boundaries}\label{stopping}
For the two selected candidate regimens carried forward from Phase~I, we conduct a parallel Phase~II evaluation, with a maximum total sample size of $N$ for each regimen, including patients previously accrued at that regimen in Phase~I. We consider $R$ interim looks and one final look (i.e., $R+1$ looks in total) at cumulative sample sizes $n_1,\ldots,n_R,N$. At each interim look, a go/no-go decision is made according to whether the futility or efficacy boundary is crossed.

Following the BOP2 design~\citep{Zhou2017-tv} and the BOP2-FE design~\citep{Xu2025-xq}, we consider efficacy and toxicity as the endpoints for dose optimization. Let $Y$ denote the joint efficacy--toxicity outcome for a patient. Specifically, $Y$ has four categories corresponding to
(1) efficacy without toxicity,
(2) efficacy with toxicity,
(3) no efficacy without toxicity, and
(4) no efficacy with toxicity.
Let $\theta_k=\Pr(Y=k)$ for $k=1,\ldots,4$, with $\theta_1+\theta_2+\theta_3+\theta_4=1$. Then, the marginal probability of efficacy is $\theta_1+\theta_2$, and the marginal probability of toxicity is $\theta_2+\theta_4$.

Suppose that, at the $r$th interim look, a total of $n_r$ patients, including those enrolled in both Phase~I and Phase~II, have been treated at a given regimen and have completed endpoint evaluation. Let $\D_{n_r}=(x_1,x_2,x_3,x_4)$ denote the combined Phase~I and Phase~II data available at that look, where $x_k$ is the number of patients with $Y=k$ and $\sum_{k=1}^4 x_k=n_r$. Assuming that $\boldsymbol{\theta}=(\theta_1,\theta_2,\theta_3,\theta_4)^T$ follows a Dirichlet prior,$(\theta_1,\theta_2,\theta_3,\theta_4)\sim\textup{Dir}(a_1,a_2,a_3,a_4)$, the posterior distribution of $\boldsymbol{\theta}$ is
$\boldsymbol{\theta}\mid \D_{n_r}\sim\textup{Dir}(a_1+x_1,a_2+x_2,a_3+x_3,a_4+x_4)$. We set $\sum_{k=1}^4 a_k=1$, which corresponds to a vague prior with prior sample size 1.

Treatment is regarded as promising only when the efficacy probability exceeds a prespecified threshold and the toxicity probability remains below a prespecified threshold. Specifically, we consider
\begin{align}
H_0:\ &\theta_1+\theta_2\le \phi_{01}
\quad \text{or} \quad
\theta_2+\theta_4\ge \phi_{02}
\nonumber\\
\text{versus}\quad
H_1:\ &\theta_1+\theta_2> \phi_{01}
\quad \text{and} \quad
\theta_2+\theta_4< \phi_{02},
\end{align}
where $\phi_{01}$ and $\phi_{02}$ are the null thresholds for efficacy and toxicity, respectively.

Let $C_F(n_r)$ and $C_E(n_r)$ denote the probability cutoffs for futility and efficacy, respectively, as functions of the interim sample size $n_r$. At the $r$th interim look, the stopping rules are as follows:
\begin{enumerate}[leftmargin=*]
\item Futility stopping: stop enrolling patients to the regimen and declare it unsuccessful if
\begin{align}
\textup{Pr}(\theta_1+\theta_2>\phi_{01}\mid \D_{n_r}) < C_F(n_r)
\quad \text{or} \quad
\textup{Pr}(\theta_2+\theta_4<\phi_{02}\mid \D_{n_r}) < C_F(n_r).
\nonumber
\end{align}

\item Efficacy stopping: stop enrolling patients to the regimen and declare it promising if
\begin{align}
\textup{Pr}(\theta_1+\theta_2>\phi_{01}\mid \D_{n_r}) \ge C_E(n_r)
\quad \text{and} \quad
\textup{Pr}(\theta_2+\theta_4<\phi_{02}\mid \D_{n_r}) \ge C_E(n_r).
\nonumber
\end{align}

\item Otherwise, continue enrolling patients until the next interim analysis.
\end{enumerate}

For the probability cutoffs for futility stopping, $C_F(n_r)$, and efficacy stopping, $C_E(n_r)$, following the BOP2-FE design~\citep{Xu2025-xq}, we use the power functions
\begin{align}
\label{eq:cf}
C_F(n_r)=\lambda \left( \frac{n_r}{N} \right)^{\gamma},
\end{align}
and
\begin{align}
\label{eq:ce}
C_E(n_r)=1-(1-\lambda) \left( \frac{n_r}{N} \right)^{\eta},
\end{align}
where $0 \le \lambda \le 1$, $0 < \gamma \le 1$, and $0 < \eta \le 3$ are tuning parameters. The parameter $\lambda$ is shared by both $C_F(n_r)$ and $C_E(n_r)$ and is chosen so that the two boundaries coincide at the final analysis when $n_{R+1}=N$. As a result, a clear final decision can always be made: at the final analysis, the treatment is declared either futile or efficacious, with no intermediate case.

To maximize statistical power while controlling the type~I error rate at a prespecified level, the tuning parameters $(\lambda, \gamma, \eta)$ are selected through the following calibration steps:
\begin{enumerate}[label=\textit{Calibration step~\Roman*:}, leftmargin=*, align=left, itemsep=0.4em]
    \item Specify $H_0$, $H_1$, the target type~I error rate, and the maximum sample size $N$.
    \item Identify the values of $(\lambda, \gamma, \eta)$ that satisfy the desired type~I error constraint, for example by grid search.
    \item Among the values identified in Calibration step~II, select the combination that yields the highest power as the optimal set of design parameters.
\end{enumerate}
Once the optimal tuning parameters are determined, the futility and efficacy stopping boundaries can be tabulated in advance and prespecified in the protocol.

\subsubsection{Operation and Final Selection}\label{design}
The objective of Phase~II is to determine whether either of the two selected candidate regimens is promising in terms of efficacy and tolerability and, if so, to identify the final recommended OBD. The two selected candidate regimens may include monotherapy, combination therapy, or both. The Phase~II evaluation is implemented through enrollment, interim monitoring, and final selection as follows:
\begin{description}[leftmargin=!, labelwidth=3.3cm]
\item[Patient enrollment.] 
For each active selected candidate regimen $j$ ($j=1,2$), enroll additional patients until the next interim look.

\item[Interim monitoring.] 
At each interim look, apply the futility and efficacy stopping boundaries defined in Section~\ref{stopping} to each active selected candidate regimen. The interim decision is made as follows:
    \begin{enumerate}[label=(\alph*), leftmargin=*]
    \item If regimen $j$ crosses the futility boundary, stop enrolling patients to that regimen and declare it unsuccessful.

    \item If regimen $j$ crosses the efficacy boundary, reject $H_0$ for that regimen and declare it promising.

    \item If both selected candidate regimens cross the efficacy boundary at the same interim look, stop the trial early and select the regimen with the larger current estimated utility as the final recommended OBD.

    \item If only one selected candidate regimen crosses the efficacy boundary, stop the trial early and select that regimen as the final recommended OBD only when its current estimated utility is larger than that of the competing selected candidate regimen.

    \item If no final recommendation is made at the interim look, continue enrolling patients in the remaining active selected candidate regimen or regimens until the next interim analysis.
    \end{enumerate}

\item[Final analysis.] 
Once the maximum sample size is reached or accrual has been stopped for both selected candidate regimens, perform the final analysis based on the complete data. The final decision is made as follows:
    \begin{enumerate}[label=(\alph*), leftmargin=*]
    \item Apply the final futility and efficacy criteria defined in Section~\ref{stopping} to each selected candidate regimen.

    \item If regimen $j$ satisfies the final efficacy criterion, declare it promising; otherwise, declare it unsuccessful.

    \item Among the candidate regimens declared promising, select the one with the larger estimated utility as the final recommended OBD.

    \item If neither candidate regimen is declared promising, no OBD is selected.
    \end{enumerate}
\end{description}

\section{Simulation studies}\label{simulation}
We conducted simulation studies to evaluate the operating characteristics of the proposed seamless staggered Phase~I/II design. The primary objectives were to assess the operating characteristics of both the proposed Phase~I procedure (including the Mono and Combo components) and the seamless Phase~II evaluation. We first describe the simulation configurations and then summarize the resulting operating characteristics.

\subsection{Simulation configurations}\label{sec:sim_config}
Agent~$A$ had $L=5$ dose levels $(A_1,\ldots,A_5)$ and agent~$B$ had $M+1=3$ levels $(B_0,\ldots,B_2)$, where $B_0$ denotes no administration of agent~$B$ (i.e., Mono with agent~$A$). Thus, the candidate regimens formed a $5\times 3$ regimen grid $\{(A_l,B_m): l=1,\ldots,5;\ m=0,1,2\}$. We considered six scenarios; for each scenario, the true toxicity probabilities $\{p_T(l,m)\}$ and treatment response probabilities $\{p_E(l,m)\}$ were specified on the $5\times 3$ grid as shown in Table~\ref{tab:tox_eff_scenarios}. 

Patients accrued sequentially at an accrual rate of $\rho=8$ patients per month. To reflect delayed outcome ascertainment, we assumed maximum assessment windows of $\tau_T=1$ month for toxicity, $\tau_S=1$ month for early activity, and $\tau_E=2$ months for treatment response. For the Phase~I BF-BOIN-ET procedure, we considered toxicity $(T)$ and a binary early activity endpoint $(S)$ for dose-assignment decisions. The early activity endpoint was defined according to prespecified criteria and was assessed earlier than the formal efficacy endpoint. The formal efficacy endpoint $(E)$ was treatment response and was used for final Phase~I candidate selection and Phase~II evaluation. The early activity probabilities for each scenario are shown in Table~\ref{tab:early_activity_scenarios}. To reflect that early activity is associated with, but not identical to, the later treatment response endpoint, the binary outcomes $(S,E)$ were generated jointly using a Gaussian copula model with Bernoulli marginal probabilities $p_S(l,m)$ and $p_E(l,m)$ and a latent correlation of 0.7.

For endpoint $X\in\{T,S,E\}$, if an event occurred, its event time was sampled from $\mathrm{Unif}(0,\tau_X)$; otherwise, the completion time was set to $\tau_X$. For the Phase~I dose-assignment endpoints $X\in\{T,S\}$, pending outcomes at each interim decision time were accommodated using the effective-data construction described in Section~\ref{methodology}.

Dose escalation in Phase~I proceeded in cohorts of size $3$. Phase~I dose-assignment decisions were based on interim estimates $(\tilde p_T,\tilde p_S)$ using two toxicity cutoffs and one early-activity cutoff: $(\lambda_{T1},\lambda_{T2})=(0.236,\,0.358)$ and $\lambda_S=0.197$. The target toxicity level (TTL) was $\phi=0.30$, and the minimum required efficacy probability for final Phase~I candidate selection was $\phi_{E0}=0.25$. Safety and early-activity monitoring used Bayesian elimination rules with independent $\mathrm{Beta}(1,1)$ priors for toxicity and early activity. Using the notation introduced in Section~\ref{methodology}, we set $\phi_{T0}=\phi=0.30$, $\phi_{S0}=p_{S,\min}=0.6\phi_{E0}=0.15$, $C_T=0.95$, and $C_S=0.90$. The elimination rules were applied as follows:
\begin{itemize}
\item \textit{Toxicity elimination:} eliminate $(l,m)$ if $\Pr\!\left\{p_T(l,m)>\phi_{T0}\mid \text{data}\right\}>C_T$.
When $(l,m)$ was eliminated for toxicity, we additionally removed from further assignment all regimens in the monotone upper-right region $\{(l',m'):\ l'\ge l,\ m'\ge m\}$.
\item \textit{Early-activity elimination:} eliminate $(l,m)$ for insufficient early activity if \\
$\Pr\!\left\{p_S(l,m)\le\phi_{S0}\mid \text{data}\right\}>C_S$.
\end{itemize}
The trial was terminated early if all candidate regimens were eliminated. Each staggered subtrial was also stopped early when the current target regimen was recommended again and the total number of patients treated at that target reached $n_{\mathrm{stop}}=12$.

At the end of Phase~I, observed toxicities were smoothed using two-dimensional isotonic regression to enforce monotonicity over the regimen grid. Let $\bar p_T(l,m)$ denote the isotonic toxicity estimate. For each fixed $B_m$ (i.e., within each column), the maximum tolerated regimen was defined as the $A_l$ whose isotonic estimate was closest to the TTL $\phi=0.30$. The admissible set consisted of non-eliminated regimens not exceeding the maximum tolerated regimen within each $B$-column. Among admissible regimens, the Phase~I utility for regimen $(A_l,B_m)$ was computed using the prespecified efficacy--toxicity utility with weights $(\psi_1,\psi_2,\psi_3,\psi_4)=(100,60,40,0)$ for the four joint outcome categories $(E=1,T=0)$, $(E=1,T=1)$, $(E=0,T=0)$, and $(E=0,T=1)$, respectively. Finally, the two candidate regimens with the highest utilities were carried forward to Phase~II. 

To the best of our knowledge, no existing designs address same objective as the proposed design. Consequently, direct comparisons to current methods would be uninterpretable or lack meaning. Therefore, we evaluated the following combinations or extensions of existing dose-finding designs, which can be viewed as naive or modified versions of the proposed method: (i) BOIN for Mono combined with the waterfall design for Combo (BOIN-W)~\citep{liu2015bayesian,Zhang2016-fp}; (ii) TITE-BOIN for Mono combined with a time-to-event implementation of the waterfall design for Combo (TITE-BOIN-W); (iii) BOIN-ET for Mono combined with BOIN-ETC for Combo (BOIN-ET)~\citep{takeda2018boin,Kakizume2024-sr}; (iv) TITE-BOIN-ET for Mono combined with a time-to-event implementation of BOIN-ETC for Combo (TITE-BOIN-ET)~\citep{Takeda2020TITEBOINET,Kakizume2024-sr}; and (v) a continual reassessment method (CRM) comparator using CRM for Mono and CRM within each one-dimensional waterfall subtrial for Combo (CRM-W). Given prior evidence suggesting comparable OBD selection performance between model-based and model-assisted approaches, we adopted combinations of model-assisted designs because of their greater implementation simplicity. Detailed specifications for each design are provided in the Supplemental Material.

For Phase~II, we evaluated the top two candidate regimens selected from Phase~I. For each candidate, the maximum total sample size was set to $N=40$, including patients accrued at that candidate in Phase~I; therefore, up to $40-n_{\mathrm{Ph1}}$ additional patients were enrolled in Phase~II, where $n_{\mathrm{Ph1}}$ is the Phase~I sample size at that candidate. Interim looks were conducted at cumulative sample sizes $n\in\{20,30,40\}$, again including the Phase~I patients at that candidate.

Let $Y$ denote the joint efficacy--toxicity outcome for a patient. In the implementation, $Y$ had $K=4$ categories: $1=(\text{response, no toxicity})$, $2=(\text{response, toxicity})$, $3=(\text{no response, no toxicity})$, and $4=(\text{no response, toxicity})$, with probabilities $(\theta_1,\theta_2,\theta_3,\theta_4)$. We used a Dirichlet--multinomial model with symmetric prior $(\theta_1,\theta_2,\theta_3,\theta_4)\sim \mathrm{Dir}(0.25,0.25,0.25,0.25)$. At each look, the marginal efficacy (ORR) and toxicity probabilities were $\Pr(E=1)=\theta_1+\theta_2$ and $\Pr(T=1)=\theta_2+\theta_4$.

The probability cutoffs (\ref{eq:cf}) and (\ref{eq:ce}) were specified by power-family functions of the cumulative sample size. These parameters were calibrated by grid search using the BOP2-FE framework to control the one-sided type~I error at $0.10$ under the null configuration $H_0=(0.15,\,0.10,\,0.45,\,0.30)$,
and to provide adequate power under the alternative configuration $H_1=(0.45,\,0.05,\,0.35,\,0.15)$.
Under these specifications, the corresponding marginal ORR and toxicity rates were $(\phi_{01},\phi_{02})=(0.25,\,0.40)$ under $H_0$, and $(\phi_{11},\phi_{12})=(0.50,\,0.20)$ under $H_1$.

At each interim look, for each active candidate regimen, posterior probabilities for efficacy and toxicity were computed from the Dirichlet posterior. The Phase~II monitoring and final selection rules followed those described in Section~\ref{design}. Specifically, early stopping and final recommendation were based on the efficacy criterion together with the posterior mean efficacy--toxicity utility. In cases of equal utility, ties were resolved in favor of the lower-dose regimen.

For each scenario and design, 10,000 simulated trials were conducted. For each dose level $B_m$, the true maximum tolerated regimen, denoted by $d_{\mathrm{MTR},m}^{\mathrm{true}}$, was defined as the regimen $A_lB_m$ whose true toxicity probability was closest to the target toxicity level $\phi=0.30$; in the event of a tie, the higher dose level of agent~$A$ was selected. The true admissible set consisted of all regimens that did not exceed $d_{\mathrm{MTR},m}^{\mathrm{true}}$ within each dose level of agent~$B$. For each regimen in the true admissible set, the true efficacy--toxicity utility was calculated using the prespecified utility weights $(\psi_1,\psi_2,\psi_3,\psi_4)=(100,60,40,0)$. The true OBD set was defined as the regimen or regimens attaining the maximum true utility within the true admissible set. A selected candidate regimen was classified as an overdose regimen if its dose level of agent~$A$ exceeded $d_{\mathrm{MTR},m}^{\mathrm{true}}$ for the corresponding dose level $B_m$.

For Phase~I, we summarized the probability that at least one of the two selected candidate regimens belonged to the true OBD set, the probability that at least one selected candidate regimen was an overdose regimen, and the mean sample size and trial duration.

\begin{table}[!t]
\centering
\caption{Key simulation settings for Phase~I and Phase~II.}
\label{tab:sim-setting}
\begin{tabularx}{0.98\textwidth}{@{}>{\raggedright\arraybackslash}p{0.32\textwidth} >{\raggedright\arraybackslash}X@{}}
\hline
\multicolumn{2}{@{}l@{}}{\textbf{Phase~I}}\\
\hline
Dose grid & $A_1$--$A_5$ and $B_0$--$B_2$ \\
Accrual rate & $\rho=8$ patients per month \\
Dose-escalation cohort size & 3 \\
Endpoints and assessment windows & Toxicity: $\tau_T=1$ month; early activity: $\tau_S=1$ month; treatment response: $\tau_E=2$ months \\
Target toxicity level & $\phi=0.30$ \\
Decision cutoffs & $(\lambda_{T1},\lambda_{T2})=(0.236,\,0.358)$ and $\lambda_S=0.197$ \\
Utility weights for final Phase~I selection & $(\psi_1,\psi_2,\psi_3,\psi_4)=(100,60,40,0)$ \\
\hline
\multicolumn{2}{@{}l@{}}{\textbf{Phase~II}}\\
\hline
Candidate regimens & Top two utility candidate regimens selected from Phase~I \\
Maximum total sample size & $N=40$ per candidate, including Phase~I patients at that candidate \\
Interim looks & $n=20,30,40$ \\
Null and alternative configurations & $H_0=(0.15,0.10,0.45,0.30)$ and $H_1=(0.45,0.05,0.35,0.15)$ \\
Calibration target & One-sided type~I error rate of 0.10 under $H_0$ \\
\hline
\end{tabularx}
\end{table}

\subsection{Simulation results}
\begin{figure}[tbp]\centering
\includegraphics[width=\linewidth]{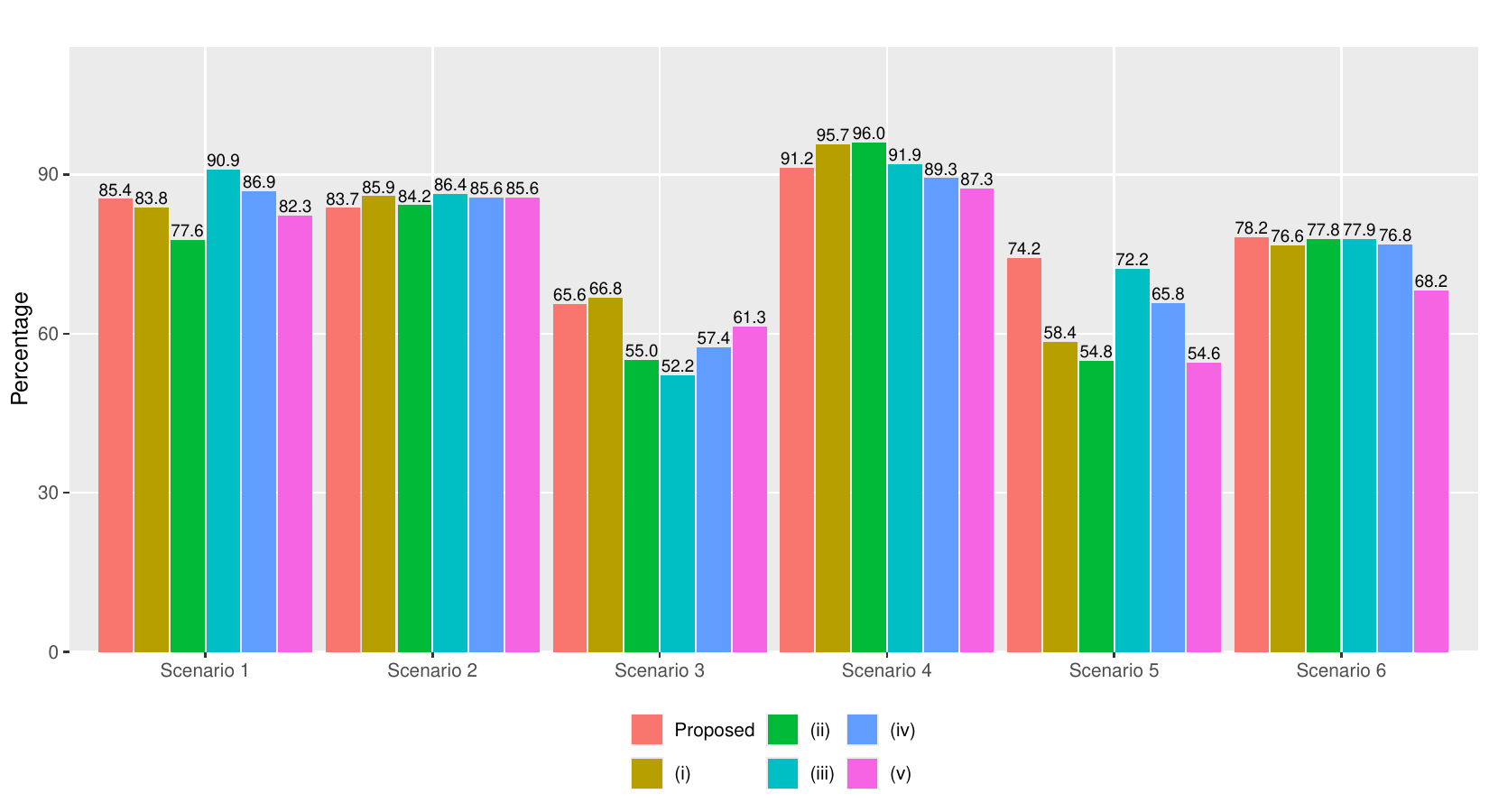}
\caption{Percentage of selecting at least one true OBD in Phase I.\\(i) BOIN-W, (ii) TITE-BOIN-W, (iii) BOIN-ET, (iv) TITE-BOIN-ET, and (v) CRM-W.}
\label{fig:fig_phaseI_correct_OBD}
\end{figure}

\begin{figure}[tbp]\centering
\includegraphics[width=\linewidth]{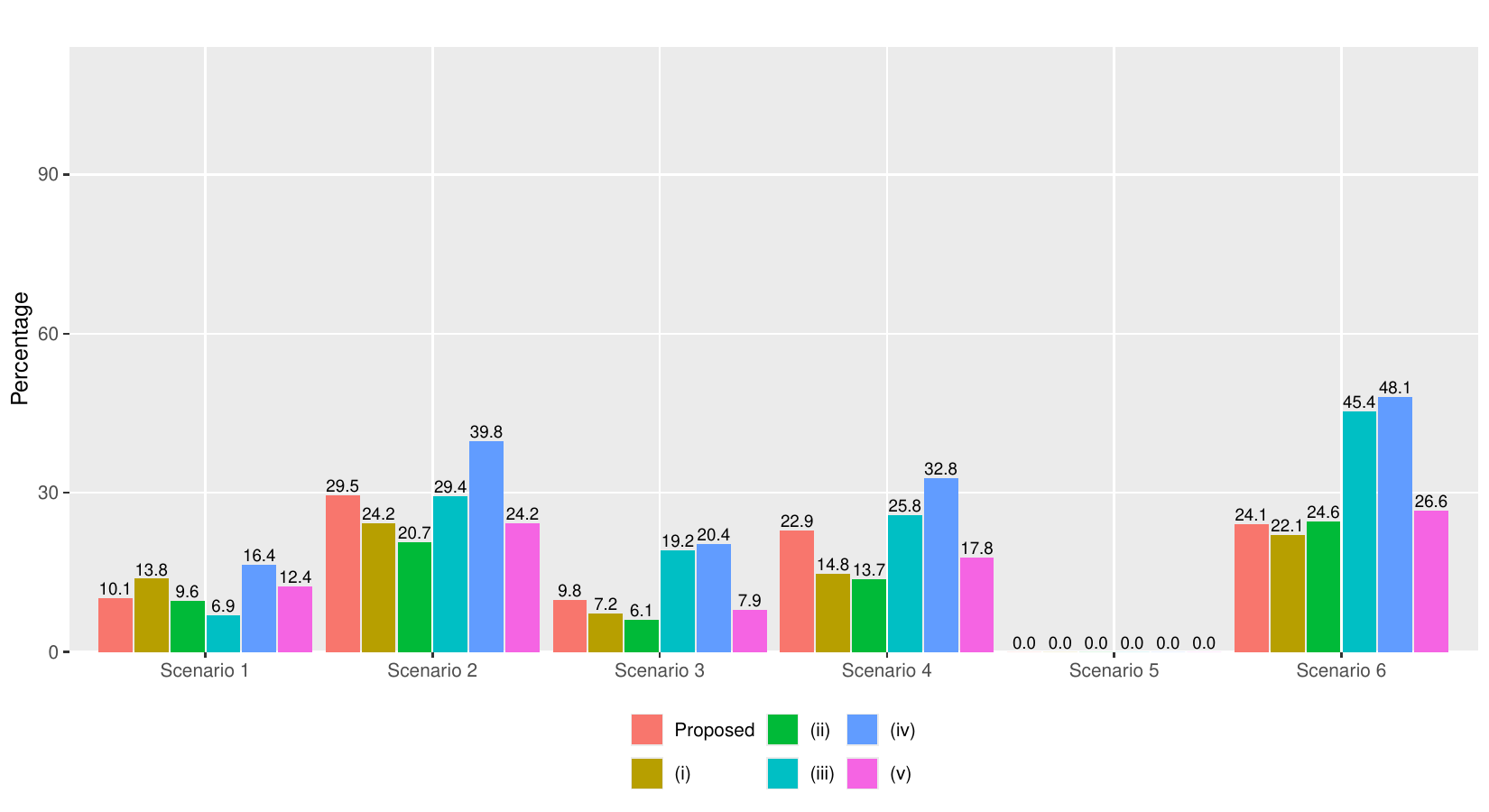}
\caption{Percentage of selecting at least one overly toxic regimen in Phase I.\\(i) BOIN-W, (ii) TITE-BOIN-W, (iii) BOIN-ET, (iv) TITE-BOIN-ET, and (v) CRM-W.}
\label{fig:fig_phaseI_overdose_regimen}
\end{figure}

\begin{figure}[tbp]\centering
\includegraphics[width=\linewidth]{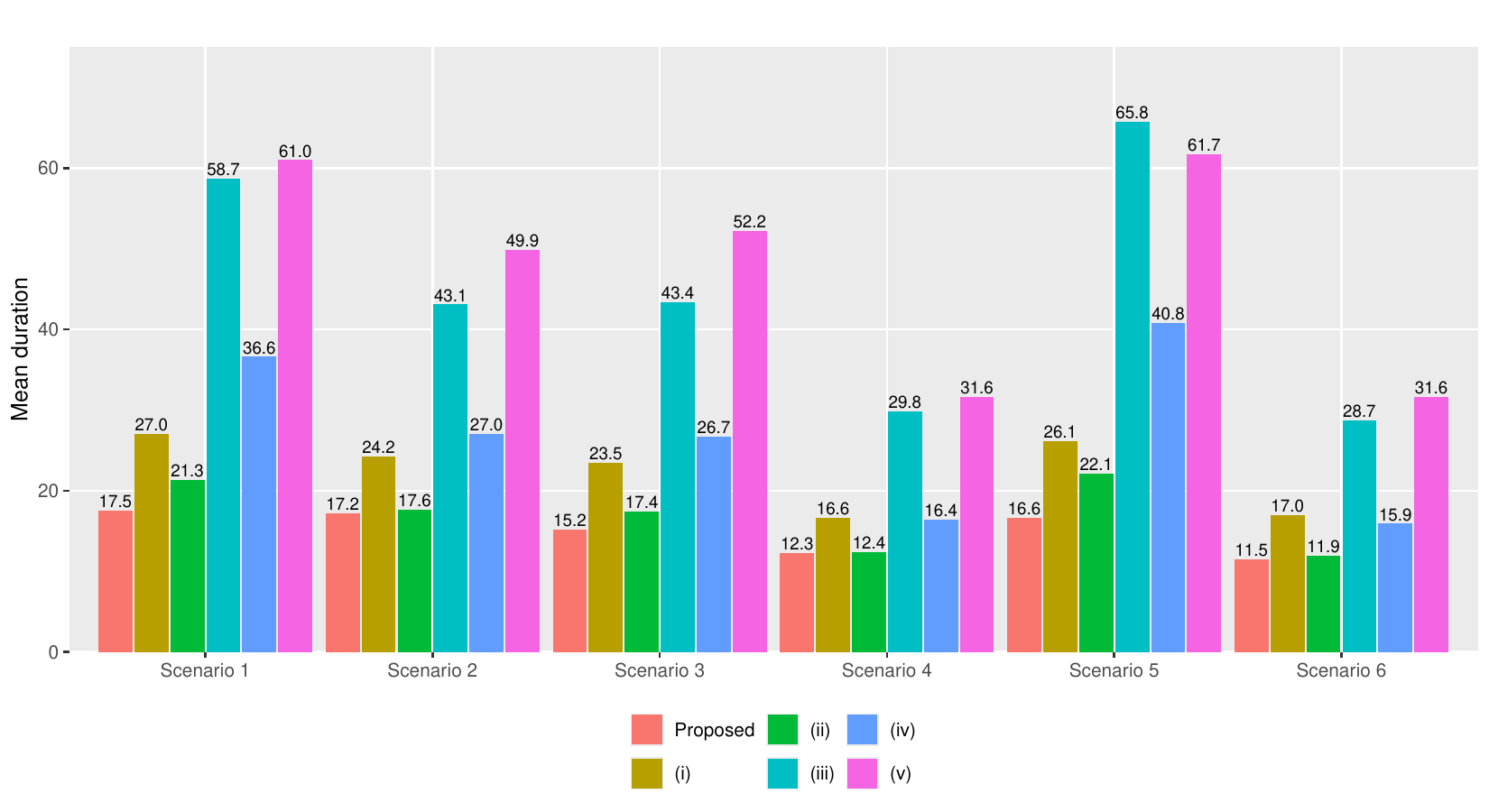}
\caption{Trial duration in Phase I.\\(i) BOIN-W, (ii) TITE-BOIN-W, (iii) BOIN-ET, (iv) TITE-BOIN-ET, and (v) CRM-W.}
\label{fig:fig_phaseI_trial_duration}
\end{figure}

\begin{figure}[tbp]\centering
\includegraphics[width=\linewidth]{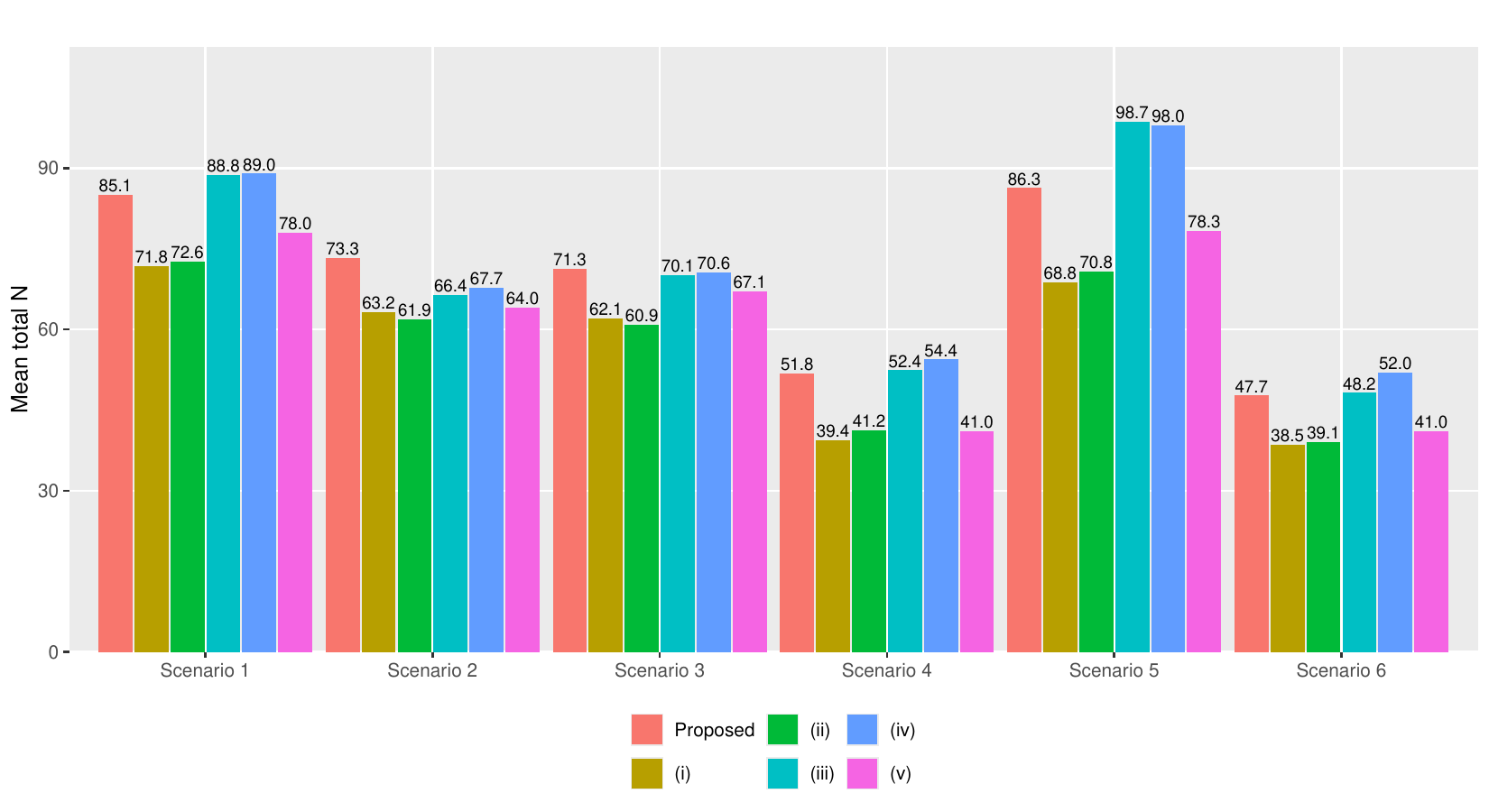}
\caption{Average number of treated patients in Phase I.\\(i) BOIN-W, (ii) TITE-BOIN-W, (iii) BOIN-ET, (iv) TITE-BOIN-ET, and (v) CRM-W.}
\label{fig:fig_phaseI_mean_totalN}
\end{figure}

First, we evaluated the operating characteristics of the seamless Mono/Combo design in Phase~I. Figure~\ref{fig:fig_phaseI_correct_OBD} shows the probability that at least one of the two selected candidate regimens was a true OBD. The proposed design achieved performance comparable to that of the conventional designs. Figure~\ref{fig:fig_phaseI_overdose_regimen} shows the probability that at least one of the two selected candidate regimens was an overly toxic regimen, which was also comparable across designs. The total trial duration is shown in Figure~\ref{fig:fig_phaseI_trial_duration}; the proposed design yielded the shortest duration in all scenarios. The mean total sample size is shown in Figure~\ref{fig:fig_phaseI_mean_totalN}, and was comparable to that of the other designs.


The Phase~II results are summarized below. Table~\ref{tab:top_summary_operating_characteristics} presents the probability of correctly selecting the final OBD in Phase~II when the two selected candidate regimens included a true OBD, together with the mean number of additional patients enrolled in Phase~II ($\bar N_{\mathrm{II}}$), the mean total sample size in Phase~I ($\bar N$), and the overall mean total sample size ($\bar N_{\mathrm{I+II}}$). The conditional probability of correctly selecting the final OBD was approximately 80\% or higher across scenarios, ranging from 79.3\% to 93.5\%. In addition, the mean number of additional patients required in Phase~II was small, ranging from 20.9 to 25.1 patients, with an average of approximately 23 patients. These results suggest that the proposed seamless Phase~II evaluation can identify the final OBD with high probability while requiring only a limited number of additional patients after Phase~I. Detailed regimen-specific futility and efficacy stopping probabilities are provided in the Supplemental Material.
\begin{table}[!htbp]
\centering
\caption{Summary of operating characteristics for the seamless Phase~I/II design}
\label{tab:top_summary_operating_characteristics}
\small
\setlength{\tabcolsep}{3.5pt}
\renewcommand{\arraystretch}{1.05}
\begin{threeparttable}
\begin{tabular}{@{}crrrrrrr@{}}
\toprule
Scenario & $P_{\mathrm{sel}}$ (\%) & $P_{\mathrm{final}}$ (\%) & $P_{\mathrm{final}\mid\mathrm{sel}}$ (\%) & $P_{\mathrm{eff}}$ (\%) & $\bar N_{\mathrm{I}}$ & $\bar N_{\mathrm{II}}$ & $\bar N_{\mathrm{I+II}}$ \\
\midrule
1 & 85.4 & 79.8 & 93.5 & 92.0 & 85.1 & 23.9 & 109.0 \\
2 & 83.7 & 77.2 & 92.3 & 83.4 & 73.3 & 25.1 & 99.2 \\
3 & 65.6 & 56.8 & 86.7 & 87.8 & 71.3 & 21.4 & 95.9 \\
4 & 91.2 & 79.4 & 87.1 & 97.9 & 51.8 & 24.5 & 78.6 \\
5 & 74.2 & 58.9 & 79.3 & 92.5 & 86.3 & 23.3 & 109.5 \\
6 & 78.2 & 65.3 & 83.4 & 91.0 & 47.7 & 20.9 & 72.4 \\
\bottomrule
\end{tabular}
\begin{tablenotes}
\footnotesize
\item $P_{\mathrm{sel}}$ denotes the probability that at least one of the two candidate regimens selected at the end of Phase~I belonged to the true OBD set. $P_{\mathrm{final}}$ denotes the probability that the final recommended OBD after Phase~II belonged to the true OBD set. $P_{\mathrm{final}\mid\mathrm{sel}}$ denotes the conditional probability of correctly selecting the final OBD in Phase~II, given that the two selected candidate regimens included a true OBD. $P_{\mathrm{eff}}$ denotes the percentage of simulations in which the selected final OBD was declared efficacious in Phase~II. $\bar N_{\mathrm{I}}$ is the mean total sample size in Phase~I, $\bar N_{\mathrm{II}}$ is the mean number of additional patients enrolled in Phase~II, and $\bar N_{\mathrm{I+II}}$ is the overall mean total sample size.
\end{tablenotes}
\end{threeparttable}
\end{table}

\section{Discussion}\label{discussion} 
We proposed and evaluated a seamless staggered phase~I/II design for dose optimization in settings where a novel agent is developed as monotherapy and in combination with an established therapy. A key feature of the proposed design is that the Combo component can be initiated in a staggered manner without waiting for definitive identification of the MTD or OBD in the Mono component, provided that accumulating Mono data show an acceptable safety profile and a prespecified combination-initiation signal. This feature is intended to mitigate the delay commonly encountered in conventional sequential Mono--Combo development, in which combination evaluation is initiated only after completion of the Mono dose-finding component. Another important feature is the seamless transition from Phase~I to Phase~II. The proposed Phase~II procedure carries forward patients treated at the two candidate regimens selected at the end of Phase~I and incorporates these data into the Phase~II evaluation. By using accumulated Phase~I information at the selected candidate regimens, the design can reduce the number of additional patients required in Phase~II while maintaining formal futility and efficacy monitoring for final OBD selection.

The proposed Phase~I framework is flexible with respect to the information used for early dose assignment. In our implementation, dose assignment was based on a model-assisted design using toxicity and an early activity endpoint. The early activity endpoint can be defined according to the clinical and biological context, including short-term PK/PD information, target engagement, biomarker evidence, or surrogate antitumor activity observed within an early treatment cycle. When no reliable early activity surrogate is available, a simpler toxicity-driven escalation strategy may be more appropriate, with backfilling used to collect additional tumor response information at tolerable and potentially promising regimens. We intentionally kept the framework operationally simple rather than introducing a complex real-time model for the joint dynamics of toxicity, activity, and efficacy. This simplicity is important for practical implementation, particularly in trials with multiple potential Combo subtrials. When several Combo subtrials are eligible to open, the current subtrial-based allocation strategy can be maintained, or a simple prioritization rule can be incorporated to focus enrollment on a limited number of promising or clinically relevant Combo subtrials. Thus, the proposed framework provides a practical and adaptable approach for accelerating Mono--Combo dose optimization while preserving the transparency of decision-making.

Simulation studies showed that, although the proposed design did not minimize the number of treated patients, it achieved the shortest Phase~I trial duration, even compared with designs incorporating TITE. The proposed design generally required more patients than BOIN-W, TITE-BOIN-W, and CRM-W, while its sample size was similar to or smaller than that of the BOIN-ET-based comparators in several scenarios. We also confirmed that the probability of selecting at least one true OBD at the end of Phase~I was comparable to that of the competing designs, while the probability of selecting an overdose regimen remained acceptable. The Phase~II results further support the efficiency of the seamless carry-forward strategy. When a true OBD was included among the two candidate regimens carried forward from Phase~I, the seamless Phase~II procedure identified the final recommended OBD with high conditional probability. This was achieved with limited additional enrollment because patients previously treated at the selected candidate regimens in Phase~I were incorporated into the Phase~II analysis and counted toward the maximum total sample size for each candidate regimen. In addition, the futility and efficacy monitoring rules could further reduce unnecessary enrollment by allowing early decisions. Across the six scenarios, the mean number of additional patients enrolled in Phase~II was approximately 21--25 patients in total across the two candidate regimens, and the selected final OBD was declared efficacious with high probability.

In conclusion, the proposed seamless staggered phase~I/II design provides a practical framework for accelerating Mono--Combo dose optimization. By allowing the Combo component to be initiated before definitive identification of the MTD or OBD in the Mono component and by carrying forward Phase~I data at the selected candidate regimens into Phase~II, the design can shorten the overall evaluation process while preserving transparent model-assisted decision-making. In the simulation studies, the proposed design shortened the Phase~I trial duration while maintaining dose-selection accuracy and safety comparable to existing approaches, and the seamless Phase~II evaluation identified the final recommended OBD with high probability using only limited additional enrollment when a true OBD was carried forward from Phase~I.

\subsection*{Acknowledgments}
For MK, this work was supported by JSPS KAKENHI (Grant Number JP26K21185).

\bibliographystyle{abbrvnat}
\bibliography{kojima}%


\begin{table}[t]
\centering
\begin{threeparttable}
\caption{True toxicity and efficacy response probabilities used in the simulation study.}
\label{tab:tox_eff_scenarios}

\small
\begin{tabular}{cc}
\toprule
\multicolumn{1}{c}{\bfseries Scenario 1} & \multicolumn{1}{c}{\bfseries Scenario 2}\\
\midrule
\begin{tabular}{lccc}
     & \textbf{B0} & \textbf{B1} & \textbf{B2} \\
\midrule
A1 & (0.01,\,0.05) & (0.05,\,0.10) & (0.07,\,0.20) \\
A2 & (0.04,\,0.10) & (0.10,\,0.20) & (0.13,\,0.30) \\
A3 & (0.11,\,0.20) & (0.15,\,0.30) & (\textbf{0.25},\,\textbf{0.60}) \\
A4 & (0.15,\,0.30) & (\textbf{0.25},\,\textbf{0.60}) & (0.45,\,0.65) \\
A5 & (\textbf{0.25},\,\textbf{0.60}) & (0.45,\,0.65) & (0.55,\,0.70) \\
\end{tabular}
&
\begin{tabular}{lccc}
     & \textbf{B0} & \textbf{B1} & \textbf{B2} \\
\midrule
A1 & (0.05,\,0.05) & (0.15,\,0.15) & (\textbf{0.25},\,\textbf{0.45}) \\
A2 & (0.10,\,0.10) & (\textbf{0.25},\,\textbf{0.45}) & (0.45,\,0.50) \\
A3 & (0.15,\,0.20) & (0.45,\,0.50) & (0.55,\,0.55) \\
A4 & (\textbf{0.25},\,\textbf{0.45}) & (0.55,\,0.55) & (0.65,\,0.60) \\
A5 & (0.45,\,0.50) & (0.65,\,0.60) & (0.75,\,0.65) \\
\end{tabular}
\\[1em]

\midrule
\multicolumn{1}{c}{\bfseries Scenario 3} & \multicolumn{1}{c}{\bfseries Scenario 4}\\
\midrule
\begin{tabular}{lccc}
     & \textbf{B0} & \textbf{B1} & \textbf{B2} \\
\midrule
A1 & (0.10,\,0.10) & (0.10,\,0.30) & (0.15,\,0.40) \\
A2 & (0.15,\,0.20) & (0.15,\,0.40) & (\textbf{0.25},\,\textbf{0.60}) \\
A3 & (0.25,\,0.30) & (\textbf{0.25},\,\textbf{0.60}) & (0.45,\,0.60) \\
A4 & (0.50,\,0.40) & (0.50,\,0.60) & (0.60,\,0.60) \\
A5 & (0.60,\,0.60) & (0.60,\,0.60) & (0.70,\,0.60) \\
\end{tabular}
&
\begin{tabular}{lccc}
     & \textbf{B0} & \textbf{B1} & \textbf{B2} \\
\midrule
A1 & (0.10,\,0.20) & (\textbf{0.30},\,\textbf{0.60}) & (0.50,\,0.60) \\
A2 & (\textbf{0.30},\,\textbf{0.60}) & (0.45,\,0.60) & (0.60,\,0.60) \\
A3 & (0.45,\,0.60) & (0.55,\,0.60) & (0.70,\,0.60) \\
A4 & (0.55,\,0.60) & (0.65,\,0.60) & (0.80,\,0.60) \\
A5 & (0.65,\,0.60) & (0.75,\,0.60) & (0.85,\,0.60) \\
\end{tabular}
\\[1em]

\midrule
\multicolumn{1}{c}{\bfseries Scenario 5} & \multicolumn{1}{c}{\bfseries Scenario 6}\\
\midrule
\begin{tabular}{lccc}
     & \textbf{B0} & \textbf{B1} & \textbf{B2} \\
\midrule
A1 & (0.01,\,0.05) & (0.03,\,0.10) & (0.05,\,0.20) \\
A2 & (0.03,\,0.10) & (0.05,\,0.20) & (0.08,\,0.30) \\
A3 & (0.05,\,0.20) & (0.08,\,0.30) & (0.10,\,0.40) \\
A4 & (0.08,\,0.30) & (0.10,\,0.40) & (0.15,\,0.50) \\
A5 & (\textbf{0.25},\,\textbf{0.60}) & (\textbf{0.25},\,\textbf{0.60}) & (\textbf{0.25},\,\textbf{0.60}) \\
\end{tabular}
&
\begin{tabular}{lccc}
     & \textbf{B0} & \textbf{B1} & \textbf{B2} \\
\midrule
A1 & (0.15,\,0.35)$^{\textcolor{red}{\ast}}$ & (\textbf{0.25},\,\textbf{0.60}) & (0.45,\,0.60) \\
A2 & (0.45,\,0.50) & (0.45,\,0.60) & (0.55,\,0.60) \\
A3 & (0.55,\,0.60) & (0.55,\,0.60) & (0.66,\,0.60) \\
A4 & (0.65,\,0.60) & (0.65,\,0.60) & (0.71,\,0.60) \\
A5 & (0.75,\,0.60) & (0.75,\,0.60) & (0.75,\,0.60) \\
\end{tabular}
\\

\bottomrule
\end{tabular}
\begin{tablenotes}
\footnotesize
\item Note: Each entry represents the true toxicity probability and tumor response probability, respectively. The tumor response probability was used as the formal efficacy endpoint for final Phase~I candidate selection and Phase~II evaluation. The probabilities at true OBDs across both Mono and Combo are shown in boldface. The red asterisk marks the dose level considered to be the OBD based on Mono alone in cases where no OBD was identified for Mono when considering the overall results.
\end{tablenotes}
\end{threeparttable}
\end{table}

\begin{table}[t]
\centering
\begin{threeparttable}
\caption{True early activity probabilities used in Phase~I simulations.}
\label{tab:early_activity_scenarios}

\small
\begin{tabular}{cc}
\toprule
\multicolumn{1}{c}{\bfseries Scenario 1} & \multicolumn{1}{c}{\bfseries Scenario 2}\\
\midrule
\begin{tabular}{lccc}
& \textbf{B0} & \textbf{B1} & \textbf{B2} \\
\midrule
A1 & 0.06 & 0.12 & 0.25 \\
A2 & 0.12 & 0.25 & 0.37 \\
A3 & 0.25 & 0.37 & \textbf{0.75} \\
A4 & 0.37 & \textbf{0.75} & 0.81 \\
A5 & \textbf{0.75} & 0.81 & 0.87 \\
\end{tabular}
&
\begin{tabular}{lccc}
& \textbf{B0} & \textbf{B1} & \textbf{B2} \\
\midrule
A1 & 0.06 & 0.18 & \textbf{0.56} \\
A2 & 0.12 & \textbf{0.56} & 0.62 \\
A3 & 0.25 & 0.62 & 0.68 \\
A4 & \textbf{0.56} & 0.68 & 0.75 \\
A5 & 0.62 & 0.75 & 0.81 \\
\end{tabular}
\\[1em]

\midrule
\multicolumn{1}{c}{\bfseries Scenario 3} & \multicolumn{1}{c}{\bfseries Scenario 4}\\
\midrule
\begin{tabular}{lccc}
& \textbf{B0} & \textbf{B1} & \textbf{B2} \\
\midrule
A1 & 0.12 & 0.37 & 0.50 \\
A2 & 0.25 & 0.50 & \textbf{0.75} \\
A3 & 0.37 & \textbf{0.75} & 0.75 \\
A4 & 0.50 & 0.75 & 0.75 \\
A5 & 0.75 & 0.75 & 0.75 \\
\end{tabular}
&
\begin{tabular}{lccc}
& \textbf{B0} & \textbf{B1} & \textbf{B2} \\
\midrule
A1 & 0.25 & \textbf{0.75} & 0.75 \\
A2 & \textbf{0.75} & 0.75 & 0.75 \\
A3 & 0.75 & 0.75 & 0.75 \\
A4 & 0.75 & 0.75 & 0.75 \\
A5 & 0.75 & 0.75 & 0.75 \\
\end{tabular}
\\[1em]

\midrule
\multicolumn{1}{c}{\bfseries Scenario 5} & \multicolumn{1}{c}{\bfseries Scenario 6}\\
\midrule
\begin{tabular}{lccc}
& \textbf{B0} & \textbf{B1} & \textbf{B2} \\
\midrule
A1 & 0.06 & 0.12 & 0.25 \\
A2 & 0.12 & 0.25 & 0.37 \\
A3 & 0.25 & 0.37 & 0.50 \\
A4 & 0.37 & 0.50 & 0.62 \\
A5 & \textbf{0.75} & \textbf{0.75} & \textbf{0.75} \\
\end{tabular}
&
\begin{tabular}{lccc}
& \textbf{B0} & \textbf{B1} & \textbf{B2} \\
\midrule
A1 & 0.43$^{\textcolor{red}{\ast}}$ & \textbf{0.75} & 0.75 \\
A2 & 0.62 & 0.75 & 0.75 \\
A3 & 0.75 & 0.75 & 0.75 \\
A4 & 0.75 & 0.75 & 0.75 \\
A5 & 0.75 & 0.75 & 0.75 \\
\end{tabular}
\\

\bottomrule
\end{tabular}
\begin{tablenotes}
\footnotesize
\item Note: Each entry represents the true early activity probability $p_S(l,m)$ used for Phase~I dose-assignment decisions. The value was obtained from the corresponding tumor response probability $p_E(l,m)$ as $p_S(l,m)=\lfloor 100(10p_E(l,m)/8)\rfloor/100$. Boldface indicates the regimens corresponding to the true OBDs. The red asterisk marks the dose level considered to be the OBD based on Mono alone.
\end{tablenotes}
\end{threeparttable}
\end{table}

\clearpage
\newcommand{\beginsupplement}{%
        \setcounter{table}{0}
        \renewcommand{\thetable}{S\arabic{table}}%
        \setcounter{figure}{0}
        \renewcommand{\thefigure}{S\arabic{figure}}%
        \setcounter{section}{0}
        \renewcommand{\thesection}{S\arabic{section}}%
     }
\beginsupplement
\setcounter{page}{1}
\begin{singlespace}

\begin{center}
   \huge \textbf{Supplementary Materials} 
\end{center}

This document contains supplemental materials to the article "A staggered seamless dose-optimization design for co-developing monotherapy and combination therapy."

\subsection{Comparator-specific simulation settings}

Only implementation details specific to the comparator designs are summarized here; all simulation scenarios, utility scores, and common performance metrics were identical to those used for the proposed design. In the notation below, ``W'' denotes the waterfall structure for combination dose finding~\citep{Zhang2016-fp}.

\paragraph{(i) BOIN-W.}
BOIN-W combined the conventional BOIN design for the Mono component~\citep{liu2015bayesian} with the waterfall design for the Combo component~\citep{Zhang2016-fp}. Thus, the Combo dose matrix was evaluated through the one-dimensional subtrial structure of the waterfall design. BOIN-W was treated as a toxicity-driven comparator.

\paragraph{(ii) TITE-BOIN-W.}
TITE-BOIN-W used the same waterfall structure as BOIN-W, but applied the time-to-event BOIN approach to accommodate pending toxicity outcomes in the Mono component and within the one-dimensional Combo subtrials.

\paragraph{(iii) BOIN-ET.}
For brevity, we use BOIN-ET to denote the comparator that combined the one-dimensional BOIN-ET design for the Mono component~\citep{takeda2018boin} with the BOIN-ETC design for the Combo component~\citep{Kakizume2024-sr}. Thus, both toxicity and efficacy were used directly for dose assignment and OBD selection in the Mono and Combo components.

\paragraph{(iv) TITE-BOIN-ET.}
TITE-BOIN-ET used the same Mono--Combo structure as BOIN-ET, with TITE-BOIN-ET applied to the Mono component~\citep{Takeda2020TITEBOINET} and a time-to-event implementation of BOIN-ETC applied to the Combo component~\citep{Kakizume2024-sr}. 

\paragraph{(v) CRM-W.}
CRM-W combined a Bayesian one-parameter empiric continual reassessment method (CRM) for the Mono component with the same CRM applied within each one-dimensional subtrial of the waterfall structure for the Combo component~\citep{Zhang2016-fp}. The target toxicity probability was 0.30, and the empiric CRM model was $p_j(\beta)=s_j^{\exp(\beta)}$ with $\beta\sim N(0,1.34)$, where $s_j$ denotes the prespecified working toxicity probability at dose $j$. Dose escalation was conducted in cohorts of three without upward dose skipping.

\subsection{Supplemental simulation results}
\begin{table}[!h]
\centering
\caption{Percentage of correct OBDC selection in Phase I (Mono and Combo)}
\label{tab:correct_obd_percent}
\begin{tabular}[t]{@{}lrrrrrr@{}}
\toprule
& \multicolumn{6}{c}{Design}\\
& Proposed & (i) & (ii) & (iii) & (iv) & (v) \\
\midrule
Scenario 1 & 85.4 & 83.8 & 77.6 & 90.9 & 86.9 & 82.3 \\
Scenario 2 & 83.7 & 85.9 & 84.2 & 86.4 & 85.6 & 85.6 \\
Scenario 3 & 65.6 & 66.8 & 55.0 & 52.2 & 57.4 & 61.3 \\
Scenario 4 & 91.2 & 95.7 & 96.0 & 91.9 & 89.3 & 87.3 \\
Scenario 5 & 74.2 & 58.4 & 54.8 & 72.2 & 65.8 & 54.6 \\
Scenario 6 & 78.2 & 76.6 & 77.8 & 77.9 & 76.8 & 68.2 \\
\bottomrule
\end{tabular}
\end{table}

\begin{table}[!h]
\centering
\caption{Percentage of overdose OBDC selection in Phase I (Mono and Combo)}
\label{tab:overdose_obd_percent}
\begin{tabular}[t]{@{}lrrrrrr@{}}
\toprule
& \multicolumn{6}{c}{Design}\\
& Proposed & (i) & (ii) & (iii) & (iv) & (v) \\
\midrule
Scenario 1 & 10.1 & 13.8 & 9.6 & 6.9 & 16.4 & 12.4 \\
Scenario 2 & 29.5 & 24.2 & 20.7 & 29.4 & 39.8 & 24.2 \\
Scenario 3 & 9.8 & 7.2 & 6.1 & 19.2 & 20.4 & 7.9 \\
Scenario 4 & 22.9 & 14.8 & 13.7 & 25.8 & 32.8 & 17.8 \\
Scenario 5 & 0.0 & 0.0 & 0.0 & 0.0 & 0.0 & 0.0 \\
Scenario 6 & 24.1 & 22.1 & 24.6 & 45.4 & 48.1 & 26.6 \\
\bottomrule
\end{tabular}
\end{table}

\begin{table}[!h]
\centering
\caption{Mean total number of patients treated in Phase I (Mono and Combo)}
\label{tab:mean_totalN}
\begin{tabular}[t]{@{}lrrrrrr@{}}
\toprule
& \multicolumn{6}{c}{Design}\\
& Proposed & (i) & (ii) & (iii) & (iv) & (v) \\
\midrule
Scenario 1 & 85.1 & 71.8 & 72.6 & 88.8 & 89.0 & 78.0 \\
Scenario 2 & 73.3 & 63.2 & 61.9 & 66.4 & 67.7 & 64.0 \\
Scenario 3 & 71.3 & 62.1 & 60.9 & 70.1 & 70.6 & 67.1 \\
Scenario 4 & 51.8 & 39.4 & 41.2 & 52.4 & 54.4 & 41.0 \\
Scenario 5 & 86.3 & 68.8 & 70.8 & 98.7 & 98.0 & 78.3 \\
Scenario 6 & 47.7 & 38.5 & 39.1 & 48.2 & 52.0 & 41.0 \\
\bottomrule
\end{tabular}
\end{table}

\begin{table}[!h]
\centering
\caption{Trial duration (mean) until last observation (Mono and Combo)}
\label{tab:trial_duration}
\begin{tabular}[t]{@{}lrrrrrr@{}}
\toprule
& \multicolumn{6}{c}{Design}\\
& Proposed & (i) & (ii) & (iii) & (iv) & (v) \\
\midrule
Scenario 1 & 17.5 & 27.0 & 21.3 & 58.7 & 36.6 & 61.0 \\
Scenario 2 & 17.2 & 24.2 & 17.6 & 43.1 & 27.0 & 49.9 \\
Scenario 3 & 15.2 & 23.5 & 17.4 & 43.4 & 26.7 & 52.2 \\
Scenario 4 & 12.3 & 16.6 & 12.4 & 29.8 & 16.4 & 31.6 \\
Scenario 5 & 16.6 & 26.1 & 22.1 & 65.8 & 40.8 & 61.7 \\
Scenario 6 & 11.5 & 17.0 & 11.9 & 28.7 & 15.9 & 31.6 \\
\bottomrule
\end{tabular}
\end{table}

\begin{table}[!htbp]
\centering
\caption{Percentages of OBD selection in Phase II}
\label{tab:pobdc_allscen}
\small
\setlength{\tabcolsep}{5pt}
\renewcommand{\arraystretch}{1.05}

\begin{tabular}{@{}l r r r | l r r r | l r r r | l r r r@{}}
\toprule
\multicolumn{4}{c}{\textbf{Scenario 1}\rule{0pt}{2.6ex}} & \multicolumn{4}{c}{\textbf{Scenario 2}\rule{0pt}{2.6ex}} & \multicolumn{4}{c}{\textbf{Scenario 3}\rule{0pt}{2.6ex}} & \multicolumn{4}{c}{\textbf{Scenario 4}\rule{0pt}{2.6ex}} \\
\addlinespace[0.4ex]
\cmidrule(lr){1-4}\cmidrule(lr){5-8}\cmidrule(lr){9-12}\cmidrule(lr){13-16}
 & B0 & B1 & B2 &  & B0 & B1 & B2 &  & B0 & B1 & B2 &  & B0 & B1 & B2 \\
\midrule
A1 & 0.0 & 0.0 & 0.3 & A1 & 0.0 & 0.2 & 23.8 & A1 & 0.0 & 4.7 & 11.2 & A1 & 1.4 & 40.3 & 0.9 \\
A2 & 0.0 & 0.5 & 2.9 & A2 & 0.0 & 25.1 & 2.3 & A2 & 1.3 & 11.5 & 32.9 & A2 & 39.1 & 1.8 & 0.0 \\
A3 & 1.1 & 2.4 & 26.1 & A3 & 1.9 & 2.1 & 0.1 & A3 & 3.1 & 23.9 & 1.1 & A3 & 1.6 & 0.0 & 0.0 \\
A4 & 4.9 & 22.2 & 1.4 & A4 & 28.4 & 0.1 & 0.0 & A4 & 0.2 & 0.1 & 0.0 & A4 & 0.0 & 0.0 & 0.0 \\
A5 & 31.5 & 1.1 & 0.0 & A5 & 1.6 & 0.0 & 0.0 & A5 & 0.0 & 0.0 & 0.0 & A5 & 0.0 & 0.0 & 0.0 \\
\bottomrule
\end{tabular}
\vspace{1.0ex}

\begin{tabular}{@{}l r r r | l r r r@{}}
\toprule
\multicolumn{4}{c}{\textbf{Scenario 5}\rule{0pt}{2.6ex}} & \multicolumn{4}{c}{\textbf{Scenario 6}\rule{0pt}{2.6ex}} \\
\addlinespace[0.4ex]
\cmidrule(lr){1-4}\cmidrule(lr){5-8}
 & B0 & B1 & B2 &  & B0 & B1 & B2 \\
\midrule
A1 & 0.0 & 0.0 & 0.2 & A1 & 15.3 & 65.3 & 2.6 \\
A2 & 0.0 & 0.1 & 1.3 & A2 & 2.4 & 1.2 & 0.0 \\
A3 & 0.8 & 2.1 & 6.1 & A3 & 0.1 & 0.0 & 0.0 \\
A4 & 5.6 & 7.4 & 14.8 & A4 & 0.0 & 0.0 & 0.0 \\
A5 & 36.7 & 14.2 & 8.0 & A5 & 0.0 & 0.0 & 0.0 \\
\bottomrule
\end{tabular}
\end{table}

\begin{table}[!htbp]
\centering
\caption{Stop for futility (\%) in Phase II}
\label{tab:futility_allscen}
\small
\setlength{\tabcolsep}{5pt}
\renewcommand{\arraystretch}{1.05}

\begin{tabular}{@{}l r r r | l r r r | l r r r | l r r r@{}}
\toprule
\multicolumn{4}{c}{\textbf{Scenario 1}\rule{0pt}{2.6ex}} &
\multicolumn{4}{c}{\textbf{Scenario 2}\rule{0pt}{2.6ex}} &
\multicolumn{4}{c}{\textbf{Scenario 3}\rule{0pt}{2.6ex}} &
\multicolumn{4}{c}{\textbf{Scenario 4}\rule{0pt}{2.6ex}} \\
\addlinespace[0.4ex]
\cmidrule(lr){1-4}\cmidrule(lr){5-8}\cmidrule(lr){9-12}\cmidrule(lr){13-16}
 & B0 & B1 & B2 &  & B0 & B1 & B2 &  & B0 & B1 & B2 &  & B0 & B1 & B2 \\
\midrule
A1 & 98.9 & 89.6 & 44.2 & A1 & 90.8 & 71.2 & 2.0 & A1 & 65.8 & 10.4 & 1.2 & A1 & 49.7 & 9.8 & 69.1 \\
A2 & 91.4 & 43.1 & 6.4 & A2 & 94.2 & 2.3 & 32.5 & A2 & 48.3 & 0.7 & 1.4 & A2 & 6.6 & 48.4 & 95.6 \\
A3 & 44.0 & 6.0 & 1.0 & A3 & 49.5 & 37.3 & 76.7 & A3 & 9.5 & 1.0 & 32.0 & A3 & 31.7 & 86.3 & 100.0 \\
A4 & 9.0 & 0.9 & 32.0 & A4 & 2.6 & 73.3 & 100.0 & A4 & 58.0 & 45.6 & 100.0 & A4 & 73.3 & 100.0 & 100.0 \\
A5 & 1.8 & 22.6 & 88.5
   & A5 & 41.5 & 100.0 & \multicolumn{1}{c|}{NA}
   & A5 & 100.0 & \multicolumn{1}{c}{NA} & \multicolumn{1}{c|}{NA}
   & A5 & 100.0 & \multicolumn{1}{c}{NA} & \multicolumn{1}{c}{NA} \\
\bottomrule
\end{tabular}

\vspace{1.0ex}

\begin{tabular}{@{}l r r r | l r r r@{}}
\toprule
\multicolumn{4}{c}{\textbf{Scenario 5}\rule{0pt}{2.6ex}} &
\multicolumn{4}{c}{\textbf{Scenario 6}\rule{0pt}{2.6ex}} \\
\addlinespace[0.4ex]
\cmidrule(lr){1-4}\cmidrule(lr){5-8}
 & B0 & B1 & B2 &  & B0 & B1 & B2 \\
\midrule
A1 & 99.1 & 91.7 & 30.8 & A1 & 8.8 & 4.0 & 48.2 \\
A2 & 88.9 & 34.8 & 3.6 & A2 & 38.6 & 48.0 & 88.4 \\
A3 & 41.3 & 6.2 & 0.6 & A3 & 68.1 & 92.3 & 100.0 \\
A4 & 11.1 & 0.2 & 0.0 & A4 & 100.0 & \multicolumn{1}{c}{NA} & \multicolumn{1}{c}{NA} \\
A5 & 2.3 & 0.7 & 0.5 & A5 & \multicolumn{1}{c}{NA} & \multicolumn{1}{c}{NA} & \multicolumn{1}{c}{NA} \\
\bottomrule
\end{tabular}
\end{table}

\begin{table}[!htbp]
\centering
\caption{Stop for efficacy (\%) in Phase II}
\label{tab:efficacy_allscen}
\small
\setlength{\tabcolsep}{5pt}
\renewcommand{\arraystretch}{1.05}

\begin{tabular}{@{}l r r r | l r r r | l r r r | l r r r@{}}
\toprule
\multicolumn{4}{c}{\textbf{Scenario 1}\rule{0pt}{2.6ex}} &
\multicolumn{4}{c}{\textbf{Scenario 2}\rule{0pt}{2.6ex}} &
\multicolumn{4}{c}{\textbf{Scenario 3}\rule{0pt}{2.6ex}} &
\multicolumn{4}{c}{\textbf{Scenario 4}\rule{0pt}{2.6ex}} \\
\addlinespace[0.4ex]
\cmidrule(lr){1-4}\cmidrule(lr){5-8}\cmidrule(lr){9-12}\cmidrule(lr){13-16}
 & B0 & B1 & B2 &  & B0 & B1 & B2 &  & B0 & B1 & B2 &  & B0 & B1 & B2 \\
\midrule
A1 & 0.0 & 0.0 & 7.3 & A1 & 0.0 & 1.6 & 63.0 & A1 & 0.0 & 25.3 & 49.4 & A1 & 2.1 & 55.1 & 5.2 \\
A2 & 0.3 & 10.4 & 38.3 & A2 & 0.0 & 62.5 & 21.2 & A2 & 5.2 & 38.9 & 72.2 & A2 & 54.9 & 13.8 & 0.0 \\
A3 & 5.7 & 24.7 & 69.2 & A3 & 7.6 & 21.9 & 5.3 & A3 & 16.6 & 72.8 & 18.2 & A3 & 16.7 & 0.0 & 0.0 \\
A4 & 16.8 & 65.8 & 21.1 & A4 & 59.5 & 5.9 & 0.0 & A4 & 5.9 & 15.8 & 0.0 & A4 & 6.7 & 0.0 & 0.0 \\
A5 & 60.8 & 24.9 & 7.7
   & A5 & 13.3 & 0.0 & \multicolumn{1}{c|}{NA}
   & A5 & 0.0 & \multicolumn{1}{c}{NA} & \multicolumn{1}{c|}{NA}
   & A5 & 0.0 & \multicolumn{1}{c}{NA} & \multicolumn{1}{c}{NA} \\
\bottomrule
\end{tabular}

\vspace{1.0ex}

\begin{tabular}{@{}l r r r | l r r r@{}}
\toprule
\multicolumn{4}{c}{\textbf{Scenario 5}\rule{0pt}{2.6ex}} &
\multicolumn{4}{c}{\textbf{Scenario 6}\rule{0pt}{2.6ex}} \\
\addlinespace[0.4ex]
\cmidrule(lr){1-4}\cmidrule(lr){5-8}
 & B0 & B1 & B2 &  & B0 & B1 & B2 \\
\midrule
A1 & 0.0 & 0.0 & 11.5 & A1 & 18.8 & 77.5 & 8.6 \\
A2 & 0.0 & 5.4 & 41.8 & A2 & 11.4 & 17.2 & 0.9 \\
A3 & 4.3 & 28.4 & 62.2 & A3 & 8.0 & 0.0 & 0.0 \\
A4 & 19.1 & 49.6 & 67.1 & A4 & 0.0 & \multicolumn{1}{c}{NA} & \multicolumn{1}{c}{NA} \\
A5 & 60.2 & 59.9 & 61.6 & A5 & \multicolumn{1}{c}{NA} & \multicolumn{1}{c}{NA} & \multicolumn{1}{c}{NA} \\
\bottomrule
\end{tabular}
\end{table}
\begin{table}[!htbp]
\centering
\caption{Average number of patients treated in Phase II (per dose $\mid$ selected in Phase I OBD2)}
\label{tab:meanN_phase2_perdose}
\small
\setlength{\tabcolsep}{5pt}
\renewcommand{\arraystretch}{1.05}

\begin{tabular}{@{}l r r r | l r r r | l r r r | l r r r@{}}
\toprule
\multicolumn{4}{c}{\textbf{Scenario 1}\rule{0pt}{2.6ex}} &
\multicolumn{4}{c}{\textbf{Scenario 2}\rule{0pt}{2.6ex}} &
\multicolumn{4}{c}{\textbf{Scenario 3}\rule{0pt}{2.6ex}} &
\multicolumn{4}{c}{\textbf{Scenario 4}\rule{0pt}{2.6ex}} \\
\addlinespace[0.4ex]
\cmidrule(lr){1-4}\cmidrule(lr){5-8}\cmidrule(lr){9-12}\cmidrule(lr){13-16}
 & B0 & B1 & B2 &  & B0 & B1 & B2 &  & B0 & B1 & B2 &  & B0 & B1 & B2 \\
\midrule
A1 & 16.0 & 15.1 & 12.6 & A1 & 13.6 & 13.1 & 11.3 & A1 & 10.3 & 13.5 & 10.3 & A1 & 7.7 & 12.9 & 13.8 \\
A2 & 14.0 & 15.9 & 12.2 & A2 & 12.7 & 8.5 & 13.2 & A2 & 13.0 & 10.2 & 10.0 & A2 & 11.6 & 14.0 & 14.1 \\
A3 & 14.7 & 12.6 & 10.1 & A3 & 14.7 & 12.9 & 14.3 & A3 & 13.1 & 6.8 & 12.6 & A3 & 13.0 & 15.7 & 13.7 \\
A4 & 15.8 & 6.6 & 13.3 & A4 & 13.9 & 12.8 & 15.9 & A4 & 15.7 & 12.5 & 15.5 & A4 & 16.7 & 15.5 & 17.0 \\
A5 & 12.8 & 11.2 & 11.5 & A5 & 17.3 & 14.4 & \multicolumn{1}{c|}{NA} & A5 & 15.9 & \multicolumn{1}{c}{NA} & \multicolumn{1}{c|}{NA} & A5 & 14.0 & \multicolumn{1}{c}{NA} & \multicolumn{1}{c}{NA} \\
\bottomrule
\end{tabular}
\vspace{1.0ex}

\begin{tabular}{@{}l r r r | l r r r@{}}
\toprule
\multicolumn{4}{c}{\textbf{Scenario 5}\rule{0pt}{2.6ex}} & \multicolumn{4}{c}{\textbf{Scenario 6}\rule{0pt}{2.6ex}} \\
\addlinespace[0.4ex]
\cmidrule(lr){1-4}\cmidrule(lr){5-8}
 & B0 & B1 & B2 &  & B0 & B1 & B2 \\
\midrule
A1 & 16.4 & 16.6 & 10.8 & A1 & 6.5 & 11.1 & 12.5 \\
A2 & 15.5 & 16.9 & 10.2 & A2 & 11.4 & 13.6 & 14.8 \\
A3 & 16.1 & 15.0 & 8.7 & A3 & 13.9 & 15.3 & 14.0 \\
A4 & 15.0 & 11.1 & 8.5 & A4 & 15.8 & \multicolumn{1}{c}{NA} & \multicolumn{1}{c}{NA} \\
A5 & 11.7 & 8.6 & 9.2 & A5 & \multicolumn{1}{c}{NA} & \multicolumn{1}{c}{NA} & \multicolumn{1}{c}{NA} \\
\bottomrule
\end{tabular}
\end{table}

\end{singlespace}

\end{document}

%% file: def.tex
\def\D{{\text{\boldmath $D$}}}

\def\nt{{\tilde n}}
\def\xt{{\tilde x}}
\def\pt{{\tilde p}}

\def\Dt{{\tilde D}}